\pdfoutput=1
\documentclass{vldb}

\usepackage[ruled,linesnumbered,lined]{algorithm2e}

\usepackage{amsmath}
\usepackage{enumitem}

\usepackage{graphicx}
\usepackage{subfigure}
\graphicspath{{./}}

\usepackage[breaklinks=true]{hyperref}
\usepackage{multirow}
\usepackage{balance}
\usepackage{comment}
\usepackage{tikz}

\usepackage{listings}
\lstdefinelanguage{scala}{
morekeywords={abstract,case,catch,class,def,%
do,else,extends,false,final,finally,%
for,if,implicit,import,match,mixin,%
new,null,object,override,package,%
private,protected,requires,return,sealed,%
super,this,throw,trait,true,try,%
type,val,var,while,with,yield},
otherkeywords={=>,<-,<\%,<:,>:,\#,@},
keywords=[3]{StaticFixed,RuntimeFixed,Variable,RecurDef,D,ITERATIONS},
sensitive=true,
morecomment=[l]{//},
morecomment=[n]{},
morestring=[b]",
morestring=[b]',
morestring=[b]"""
}

\usepackage{color}
\definecolor{dkgreen}{rgb}{0,0.6,0}
\definecolor{gray}{rgb}{0.5,0.5,0.5}
\definecolor{mauve}{rgb}{0.58,0,0.82}
\newcommand{\tabincell}[2]{\begin{tabular}{@{}#1@{}}#2\end{tabular}}

\begin{document}


\title{Lifetime-Based Memory Management for Distributed Data Processing Systems}

\author{
\alignauthor{\fontsize{10pt}{12pt}\selectfont Lu Lu\hspace{0.4mm}\footnotemark[2]\hspace{0.5mm}, Xuanhua Shi\hspace{0.4mm}\footnotemark[2]\hspace{0.5mm}, Yongluan Zhou\hspace{0.4mm}\footnotemark[3]\hspace{0.5mm}, Xiong Zhang\hspace{0.4mm}\footnotemark[2]\hspace{0.5mm}, Hai Jin\hspace{0.4mm}\footnotemark[2]\hspace{0.5mm}, Cheng Pei\hspace{0.4mm}\footnotemark[2]\hspace{0.5mm}, Ligang He\hspace{0.4mm}\footnotemark[4]\hspace{0.5mm}, Yuanzhen Geng\hspace{0.4mm}\footnotemark[2]}\\
       \affaddr{\fontsize{9pt}{12pt}\selectfont \footnotemark[2]\hspace{1.5mm}SCTS/CGCL, Huazhong University of Science and Technology, Wuhan, China}\\
       \affaddr{\fontsize{9pt}{12pt}\selectfont \footnotemark[3]\hspace{1.5mm}University of Southern Denmark, Denmark\hspace{25mm}\footnotemark[4]\hspace{1.5mm}University of Warwick, UK}\\
       \affaddr{\fontsize{9pt}{12pt}\selectfont \footnotemark[2]\hspace{1.5mm}\{llu,xhshi,hjin,wxzhang,cpei,yzgeng\}@hust.edu.cn,\hspace{2mm}\footnotemark[3]\hspace{1.5mm}zhou@imada.sdu.dk,\hspace{2mm}\footnotemark[4]\hspace{1.5mm}liganghe@dcs.warwick.ac.uk}
}

\maketitle

\begin{abstract}

In-memory caching of intermediate data and eager combining of data in 
shuffle buffers have been shown to be very effective in minimizing the re-computation
and I/O cost in distributed data processing systems like Spark and Flink. However, it has
also been widely reported that these techniques would create a large amount of
long-living data objects in the heap, which may quickly saturate the garbage
collector, especially when handling a large dataset, and hence would limit the
scalability of the system. To eliminate this problem, we propose a
lifetime-based memory management framework, which, by automatically analyzing
the user-\linebreak defined functions and data types, obtains the expected lifetime of the
data objects, and then allocates and releases memory space accordingly to minimize
the garbage collection overhead. In particular, we present Deca, a concrete
implementation of our proposal on top of Spark, which transparently decomposes and groups
objects with similar lifetimes into byte arrays and releases their space
altogether when their lifetimes come to an end. An extensive experimental study
using both synthetic and real datasets shows that, in comparing to Spark, Deca
is able to 1) reduce the garbage collection time by up to 99.9\%,
2) to achieve up to 22.7x speed up in terms of execution time in cases without
data spilling and 41.6x speedup in cases with data spilling, and 3) to consume
up to 46.6\% less memory.

\end{abstract}

\section{Introduction}

Distributed data processing systems, such as Spark~\cite{zaharia:spark}, process huge
volumes of data in a scale-out fashion. Unlike traditional database systems using
declarative query languages and relational (or multidimensional) data models,
these systems allow users to implement application logics through \textit{User
Defined Functions} (UDFs) and \textit{User Defined Types} (UDTs) using
high-level imperative languages (such as Java, Scala and C\# etc.), which can
then be automatically parallelized onto a large-scale cluster.

Existing researches in these systems mostly focus on scalability and
fault-tolerance issues in a distributed
environment~\cite{zaharia:late, zaharia:delay, isard:quincy, anan:mantri}.
However some recent studies~\cite{anderson:efficiency, mcsherry:cost} suggest
that the execution efficiency of individual tasks in these systems is low. A major
reason is that both the execution frameworks and user programs of these systems are implemented
using high-level imperative languages running in managed runtime platforms (such
as JVM, .NET CLR, etc.). These managed runtime platforms commonly have
built-in automatic memory management, which brings significant memory and CPU
overheads. For example, the modern tracing-based garbage collectors (GC) may
consume a large amount of CPU cycles to trace living objects in the
heap~\cite{jones:gc, bu:bloat, www:tungsten}. 

Furthermore, to improve the performance of
multi-stage and iterative computations, recently developed systems support caching of
intermediate data in the main memory~\cite{power:piccolo, shinnar:m3r,
zaharia:spark, zhang:survey} and exploit eager combining and
aggregating of data in the shuffling phases~\cite{li:onepass, shi:titan}. These
techniques would generate massive \emph{long-living} data objects in the heap,
which usually stay in the memory for a significant portion of the job execution
time. However, the unnecessary continuous tracing and marking of such large
amount of long-living objects by the GC would consume
significant CPU cycles.

In this paper, we argue that distributed data processing systems like Spark, should
employ a lifetime-based memory manager, which allocates and releases memory according
to the lifetimes of the data objects rather than relying on a conventional
tracing-based GC. To verify this concept, we present Deca, an
automatic Spark optimizer, which adopts a lifetime-based memory management
scheme for efficiently reclaiming memory space. Deca automatically analyzes the
lifetimes of objects in different data containers in Spark, such as UDF
variables, cached data blocks and shuffle buffers, and then transparently
decomposes and stores a massive number of objects with similar lifetimes into a
few number of byte arrays. In this way, the massive objects essentially bypass
the continuous tracing of the GC and their space can be released by the
destruction of the byte arrays. 

Last but not the least, Deca automatically transforms the user programs so that
the new memory layout is transparent to the users. By using the aforementioned
techniques, Deca significantly optimizes the efficiency of Spark's memory
management and at the same time keeps the generality and expressibility provided
in Spark's programming model. 
In summary, the main contributions of this paper include:

\begin{itemize}[itemsep=1pt, topsep=1pt]

\item We propose a lifetime-based memory management\linebreak scheme for distributed data
  processing systems and implement a prototype on top of Spark, which is
  able to minimize the GC overhead and eliminate the memory bloat problems in Spark.

\item We design a method that changes the in-memory representation of the object
  graph of each data item by discarding all the reference values. The raw data of
  the fields of primitive types in the object graph will be compactly stored as
  a byte sequence.

\item We propose techniques to group byte sequences of data items with the same
  lifetime into a few byte arrays, thereby simplifying space reclamation.
  Deca automatically validates the memory safety of data accessing based on
  analysis of memory usage of UDT objects. 

\item We conduct extensive evaluation on various Spark programs using both
  synthetic and real datasets. The experimental results demonstrate the
  superiority of our approach by comparing with existing methods.

\end{itemize}

\section{Overview of Deca}
\label{sec:deca}

\subsection{Java GC}
\label{subsec:gc}

In typical JVM implementations, a garbage collector (GC) attempts to reclaim
memory occupied by objects that will no longer be used. A tracing GC traces
which objects are reachable by a sequence of references from some root objects.
The unreachable ones, which are called garbages, can be reclaimed. 
Oracle's Hotspot JVM implements three GC algorithms. The default Parallel
Scavenge (PS) algorithm suspends the application and spawns several parallel GC
threads to achieve high throughput. The other two algorithms, namely Concurrent
Mark-Sweep (CMS) and Garbage-First (G1), attempt to reduce GC latency by
spawning concurrent GC threads that run simultaneously with the application
thread.

All the above collectors segregate objects into multiple generations: young
generation containing recently-allocated objects, old generation containing
older objects, and permanent generation containing class metadata. Based on the
assumption that most objects would soon become garbages, a minor GC, which only
attempts to reclaim garbages in the young generation, can be run to reclaim
enough memory space. However, if there are too many old objects, then a full (or
major) GC would be run to reclaim space occupied by the old objects. Usually, a
full GC is much more expensive than a minor GC.

\subsection{Motivating Example}
\label{subsec:examples}

\begin{figure}
\centering
\begin{lstlisting}
class DenseVector[V](val data: Array[V], 
    val offset: Int, 
    val stride: Int, 
    val length: Int) extends Vector[V] {
  def this(data: Array[V]) = 
    this(data, 0, 1, data.length)
  ...
}
class LabeledPoint(var label: Double,
    var features: Vector[Double])

val lines = sparkContext.textFile(inputPath)
val points = lines.map(line => {
  val features = new Array[Double](D)
  ...
  new LabeledPoint(new DenseVector(features), label)
}).cache()
var weights = 
  DenseVector.fill(D){2 * rand.nextDouble - 1}
for (i <- 1 to ITERATIONS) {
  val gradient = points.map { p =>
    p.features * (1 / (1 + 
        exp(-p.label * weights.dot(p.features))) - 
        1) * p.label
  }.reduce(_ + _)
  weights -= gradient
}
\end{lstlisting}
\caption{Demo Spark program of Scala LR.}
\label{code:lr}
\end{figure}

A major
concept of Spark is \emph{Resilient Distributed Dataset} (RDD), which is a
fault-tolerant dataset that can be processed in parallel by a set of UDF
operations.

We use \textit{Logistic Regression} (LR) as an example to motivate and
illustrate the optimization techniques adopted in Deca.  It is a classifier that
attempts to find an optimal hyperplane that separates the data points in a
multi-dimensional feature space into two sets. Figure~\ref{code:lr} shows the
code of LR in Spark. The raw dataset is a text file with each line containing
one data point.  Hence the first UDF is a {\ttfamily \small map} function, which
extracts the data points and store them into a set of \texttt{DenseVector}
objects and (lines 12--16). An additional \texttt{LabeledPoint} object is
created for each data point to package its feature vector and label value
together. 

To eliminate disk I/O for the subsequent iterative computation, LR uses the
{\ttfamily \small cache} operation to cache the resulting \texttt{LabeledPoint}
objects in the memory. For a large input dataset, this cache may contain a
massive number of objects. 

After generating a random separating plane (lines 18-19), it iteratively runs
another {\ttfamily \small map} function and a {\ttfamily \small reduce}
function to calculate a new gradient (lines 20--26).  Here each call of this
{\ttfamily \small map} function will create a new \texttt{DenseVector} object.
These intermediate objects will not be used any more after executing the
{\ttfamily \small reduce} function. Therefore, if the aforementioned cached data
leaves little space in the memory, then GC will be frequently run to reclaim the
space occupied by the intermediate objects and make space for newly generated
ones.
Note that, after running a number of minor GCs, JVM would run a full GC to
reclaim spaces occupied by the old objects.  However, such highly expensive
full GCs would be nearly useless because most cached data objects should not be
removed from the memory.

\subsection{Life-time based Memory Management}

\begin{figure}[t]
\centering
\includegraphics[width=0.4\textwidth]{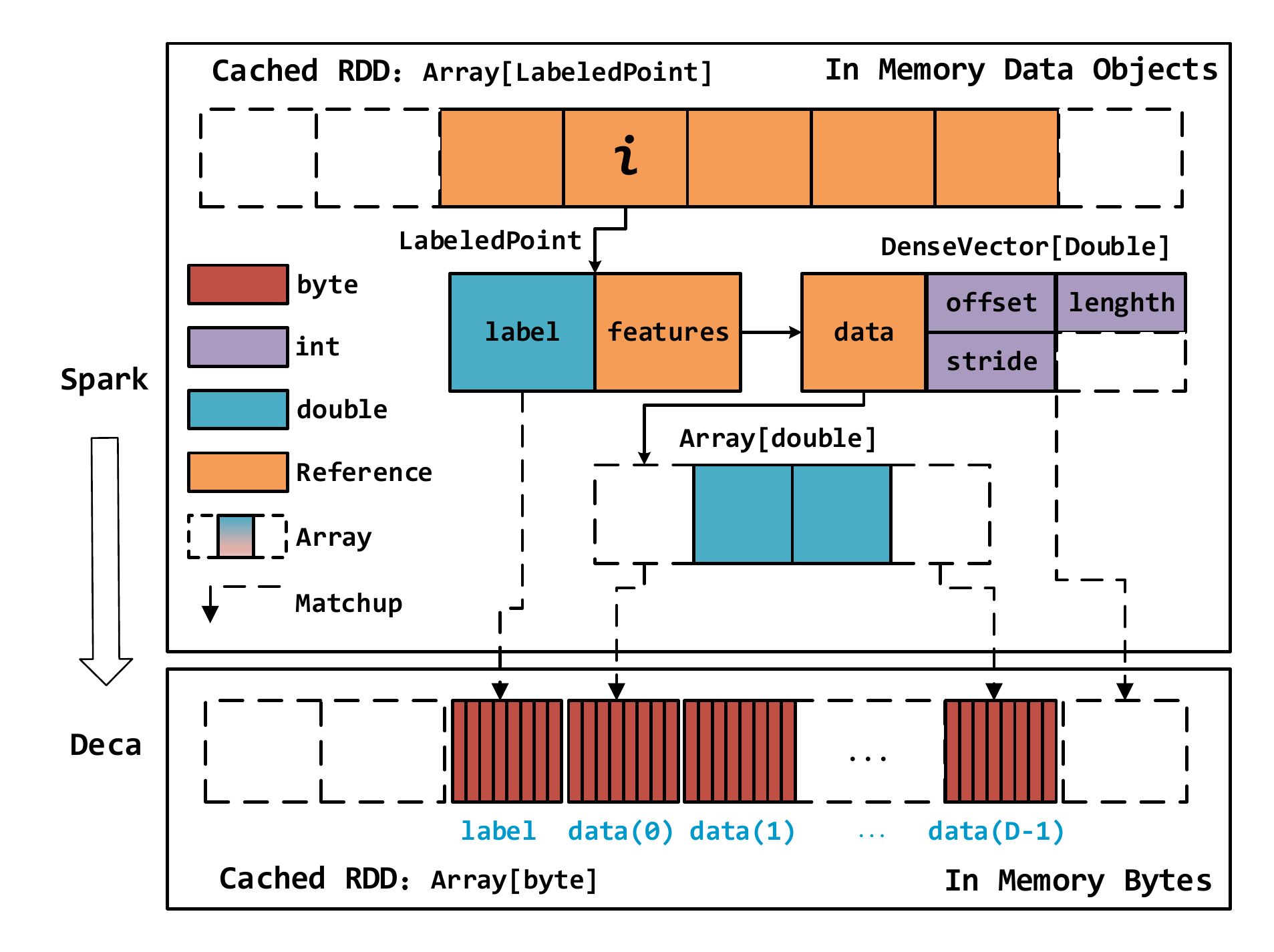}
\caption{The LR cached RDD data layout.
}
\label{fig:layout-lr}
\end{figure}

We implement the prototype of Deca based on Spark. 
In Deca, objects are stored in three types of data containers: UDF variables,
cached RDDs and Shuffle buffers. In each data container, Deca allocates a number of
fixed-sized byte arrays. 
By using the \textit{points-to} analysis~\cite{lhot:spark}, we map the UDT objects with
their appropriate containers. UDT objects are then stored in the byte arrays
after eliminating the unnecessary object headers and object references. This
compact layout would not only minimize the memory consumption of data objects
but also dramatically reduce the overhead of GC, because GC only needs
to trace a few byte arrays instead of a huge number of UDT objects. 
One can see that the size of each byte array should not be too small or too
large, otherwise it would incur high GC overheads or large unused memory spaces. 

As an example, the \texttt{LabeledPoint} objects in the LR program can be
transformed into byte arrays as shown in Figure~\ref{fig:layout-lr}. Here, all the
reference variables (in orange color, such as \texttt{features} and
\texttt{data}), as well as the headers of all the objects are eliminated. All the
cached \texttt{LabeledPoint} objects are stored into byte arrays.

The challenge of employing such a compact layout is that the space allocated to each object is
fixed. Therefore, we have to ensure that the size of an object would not exceed
its allocated space during execution so that it will not damage the data layout. This is easy for
some types of fields, such as primitive types, but less obvious
for others.  Code analysis is necessary to
identify the change patterns of the objects' sizes. Such an analysis may have a
global scope. For example, a global code analysis may identify that the
\texttt{features} arrays of all the \texttt{LabeledPoint} objects (created in
line 14 in Figure~\ref{code:lr}) actually have the same fixed size \texttt{D},
which is a global constant. Furthermore, the \texttt{features} field of a
\texttt{LabeledPoint} object is only assigned in the \texttt{LabeledPoint}
constructor. Therefore, all the \texttt{LabeledPoint} objects actually have the
same fixed size. Another interesting pattern is that in Spark applications,
objects in cached RDDs or shuffle buffers are often generated sequentially and
their sizes will not be changed once they are completely generated. Identifying
such useful patterns by a sophisticated code analysis is necessary to ensure the
safety of decomposing UDT objects and storing them compactly into byte arrays.
 
As mentioned earlier, during the execution of programs in a system like Spark,
the lifetimes of data containers created by the framework can be pre-determined
explicitly. For example, the lifetimes of objects in a cached RDD is determined
by the invocations of \texttt{cache()} and \texttt{unpersist()} in the program.
Recall that the UDT objects stored in the compact byte arrays would bypass the
GC. We put the UDT objects with the same lifetime into the same container.
For example, the cached \texttt{LabeledPoint} objects in LR have the same
lifetime, so they are stored in the same container. When a container's lifetime
comes to an end, we simply release all the references of the byte arrays in the
container, then the GC can reclaim the whole space occupied by the massive
amount of objects.

Lastly, Deca modifies the application code by replacing the codes
of object creation, field access and UDT methods with new codes that
directly write and read the byte arrays.
\section{UDT Classification Analysis}
\label{sec:classify}

\subsection{Data-size and Size-type of Objects}
To allocate enough memory space for objects, we have to estimate the object
sizes and their change patterns during runtime. Due to the complexity of object
models, to accurately estimate the size of a UDT, we have to dynamically 
traverse the runtime object reference graph of each target object and compute the
total memory consumption. Such a dynamic analysis is too costly at runtime,
especially with a large number of objects. Therefore we opt for static analysis
which only uses static object reference graphs and would not incur any runtime
overhead. We define the \textit{data-size} of an object to be the sum of the sizes
of the primitive-type fields in its static object reference graph. 
An object's \textit{data-size} is only an upper bound of the actual memory
consumption of its raw data, if one considers the cases with object sharing.

To see if UDT objects can be safely decomposed into byte
sequences, we should examine how their data-sizes change
during runtime. There are two types of UDTs that can meet the safety
requirement: 1) the data-sizes of all the instances of the UDT are
identical and do not change during runtime; or 2) the data-sizes of all
the instances of the UDT do not change during runtime.  We call these two kinds
of UDTs as \textbf{Static Fixed-Sized Type (SFST)} and \textbf{Runtime
Fixed-Sized Type (RFST)} respectively.

In addition, we call UDTs that have type-dependency cycles in their type
definition graphs as \textbf{Recursively-Defined Type}. Even without object
sharing, the instances of these types can have reference cycles in their object
graphs. Therefore, they cannot be safely decomposed. Furthermore, any UDT that
does not belong to any of the aforementioned types is called a
\textbf{Variable-Sized Type (VST)}. Once a VST object is constructed,
its data-size may change due to field assignments and method invocations during
runtime.

The objective of the UDT classification analysis is to generate the
\textbf{Size-Type} of each target UDT according to the above definitions. As 
demonstrated in Figure~\ref{fig:layout-lr}, Deca decomposes a set of objects into
primitive values and store them contiguously into compact byte sequences in a
byte array. A safe decomposition requires that the original UDT objects are
either of an SFST or an RFST. Otherwise the operations that expand the
byte sequences occupied by an object may overwrite the data of the subsequent objects in the
same byte array. 
Furthermore, as we will discuss later, an SFST can be safely decomposed in more
cases than an RFST. On the other hand, objects that do not belong to an SFST or
an RFST will not be decomposed into byte sequences in Deca. Apparently, to maximize the
effect of our approach, we should avoid overestimating the variability of the
data-size of the UDTs, which is the design goal of our following algorithms.

\subsection{Local Classification Analysis}

\begin{figure}[t]
\centering
\includegraphics[width=0.4\textwidth]{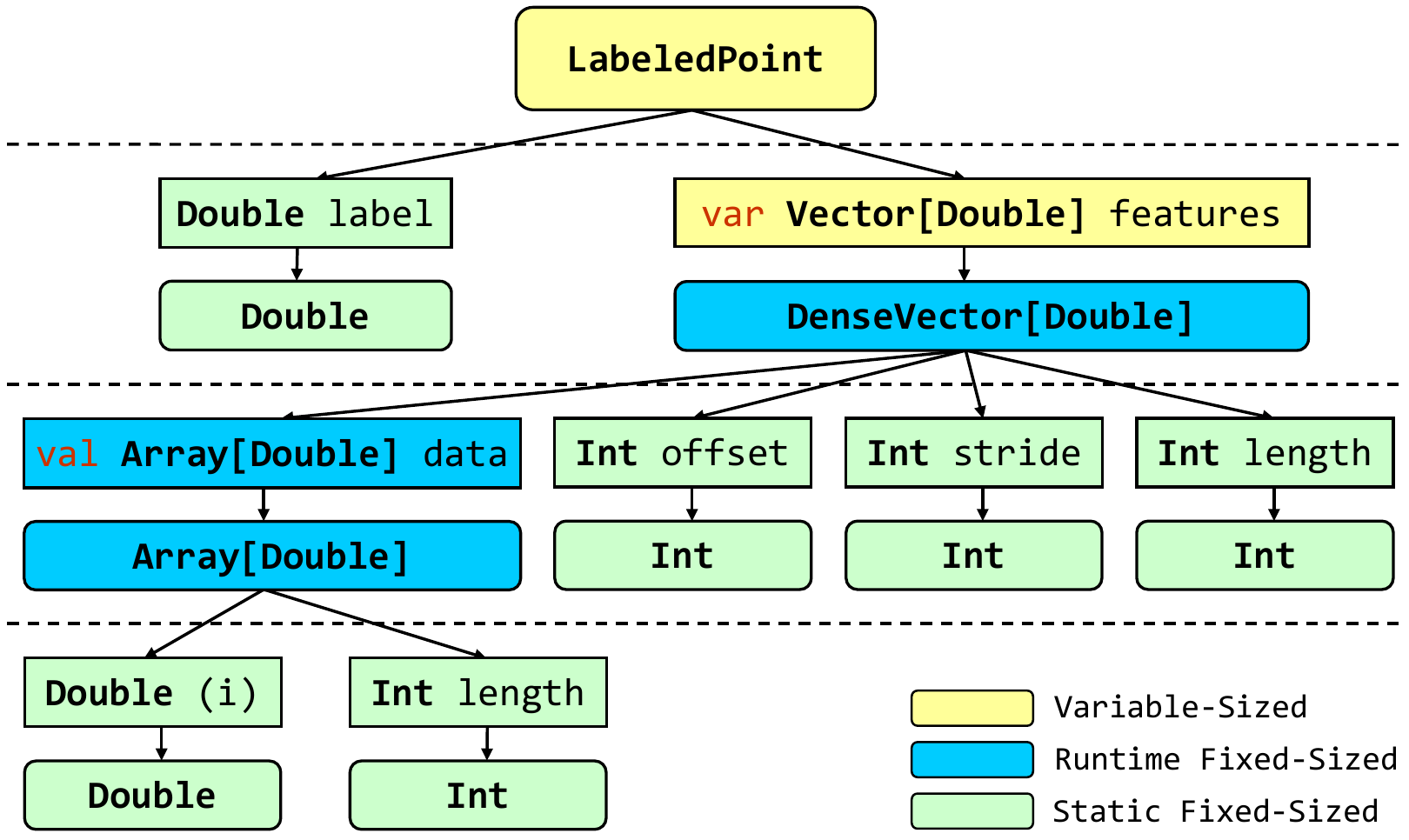}
\caption{An example of the local classification.}
\label{fig:example}
\end{figure}

The local classification algorithm analyzes an UDT by recursively traversing its
type dependency graph. For example, Figure~\ref{fig:example} illustrates the
type dependency graph of \texttt{LabeledPoint}. The size-type of
\texttt{LabeledPoint} can be determined based on the size-type of each of its
fields.

\IncMargin{0.4em}
\SetAlFnt{\small}
\begin{algorithm}[!t]

\SetKwInOut{Input}{Input}\SetKwInOut{Output}{Output}
\SetKwData{SF}{\small{StaticFixed}}\SetKwData{RF}{\small{RuntimeFixed}}
\SetKwData{V}{\small{Variable}}\SetKwData{RD}{\small{RecurDef}}
\SetKwProg{Fn}{Function}{}{end}
\SetKwFunction{AT}{AnalyzeType}\SetKwFunction{AF}{AnalyzeField}
\SetKwData{F}{\small{final}}

\Input{The top-level annotated type $T$\;}
\Output{The size-type of $T$\;}
build the type dependency graph $G$ for $T$\;
\lIf{$G$ contains the circle path}{\KwRet{\RD}}
\lElse{\KwRet{\AT{$T$}}}
\BlankLine
\Fn{\AT{$t_{arg}$}}{
  \lIf{$t_{arg}$ is a primitive type}{\KwRet{\SF}}
  \uElseIf{$t_{arg}$ is an array type}{
    $f_{e} \leftarrow$ array element field of $t_{arg}$\;
    \uIf{\AF{$f_{e}$} $=$ \SF}{
      \KwRet{\RF}\;
    }\lElse{\KwRet{\V}}
  }
  \Else{
    $result \leftarrow$ \SF\;
    \ForEach{field $f$ of type $t_{arg}$}{
      $tmp \leftarrow$ \AF{$f$}\;
      \lIf{$tmp=$ \V}{\KwRet{\V}}
      \ElseIf{$tmp=$ \RF}{
        $result \leftarrow$ \RF\;
      }
    }
    \KwRet{$result$}\;
  }
}
\BlankLine
\Fn{\AF{$f_{arg}$}}{
  $result \leftarrow$ \SF\;
  \ForEach{runtime type $t$ in $f_{arg}.getTypeSet$}{
    $tmp \leftarrow$ \AT{$t$}\;
    \lIf{$tmp=$ \V}{\KwRet{\V}}
    \ElseIf{$tmp=$ \RF}{
      \lIf{$f_{arg}$ is not \F}{\KwRet{\V}}
      \lElse{$result \leftarrow$ \RF}
    }
  }
  \KwRet{$result$}\;
}
\caption{Local classification analysis.}
\label{code:la}
\end{algorithm}

Algorithm~\ref{code:la} shows the procedure of the local classification
analysis. The input of the algorithm is an \textit{annotated type} that
contains the information of fields and methods of the target UDT. Because the
objects referenced by a field can be of any subtype of its declared type, we use
a \textit{type-set} to store all the possible runtime types of each field. The
type-set of each field is obtained in a pre-processing phase of Deca by using 
the points-to analysis~\cite{lhot:spark} (see Section~\ref{sec:impl}).

In lines 1--2, the algorithm first determines whether the target UDT is a
recursively-defined type. It builds the type dependent graph and searches for 
cycles in the graph. If a cycle is found, the algorithm immediately
returns recursively-defined type as the final result. 

Two indirect-recursive functions, AnalyzeType (lines 4--22) and AnalyzeField (lines
23--34), are used to further determine the size-type of the target UDT. The stop
condition of the recursion is when the current type is a primitive
type (line 5). We treat each array type as having a length field and an
element field. Since different instances of an array type can have different
lengths, arrays with static fixed-sized elements will be considered as an
RFST (lines 8--9). 

We define a total ordering of the variability of the
size-types (except the recursively-defined type) as follows:\linebreak $SFST <
RFST < VST$. Based on this order, the size-type of each UDT
is determined by its field that has the highest variability (lines
12--20). Furthermore, each field's final size-type is determined by the type
with the highest variability in its type-set. But a non-final field of an RFST
will be finally classified as VST, because the same field can possibly point to
objects with different data-sizes (lines 28-29).  Consider that whenever we find
a VST field, the top-level UDT must also be classified as a VST. In this case,
the function can immediately returns without further traversing the graph.

We take the type \texttt{LabeledPoint} in Figure~\ref{code:lr} as a running
example. In Figure~\ref{fig:example}, every field has a type-set with a single
element and the declared type of each filed is equal to its corresponding
runtime type except that the
{\ttfamily \small features} field has a declared type (\texttt{Vector}),
while its runtime type is \texttt{DenseVector}. Moreover, for a more
sophisticated implementation of logistic regression with high-dimensional data
sets, the {\ttfamily \small features} field can have both \texttt{DenseVector} and
\linebreak \texttt{SparseVector} in its type-set.

Since there is no cycle in the type dependency graph, \texttt{LabeledPoint} is
not a recursively-defined type. 
As shown in Figure~\ref{fig:example}, \texttt{LabeledPoint} contains a primitive
field (i.e. {\ttfamily \small label}) and a field of the \texttt{Vector} type 
( i.e. {\ttfamily \small features}). Therefore, the \textit{size-type} of
\texttt{LabeledPoint} is determined by the \textit{size-type} of {\ttfamily
\small features}, i.e. the size-type of \texttt{DenseVector}.  It contains
four fields: one of the array type and other three of the primitive type.  The
{\ttfamily \small data} field will be classified as an RFST
but not a VST due to its {\ttfamily \small final} modifier ({\ttfamily \small
val} in Scala). Furthermore, the \texttt{DenseVector} objects assigned to {\ttfamily
\small features} can have different \textit{data-size} values because they
may contain different arrays. Therefore, both {\ttfamily \small features} and
\texttt{LabeledPoint} belong to VST.

\subsection{Global Classification Analysis}

The local classification algorithm is easy to implement and has negligible
computational overhead. But it is conservative and often overestimates the
variability of the target UDT. For example, 
the local classifier conservatively assumes that the {\ttfamily \small features}
field of a \texttt{LabeledPoint} object may be assigned with \texttt{DenseVector}
objects with different \textit{data-size} values. Therefore it mistakenly
classifies it as a VST, which can not be safely decomposed.

Furthermore, the local classifier assumes that the \linebreak \texttt{DenseVector} objects
contain arrays ({\ttfamily \small features.data}) with different lengths. Even
if we change the modifier of {\ttfamily \small features} from {\ttfamily \small var} to
{\ttfamily \small val}, i.e, only allowing it to be assigned once, the local
classifier still considers it as an RFST rather than an SFST. 

For UDTs that are categorized as RFST or VST, we further propose an algorithm to
refine the classification results via global code analysis on the relevant
methods of the UDTs.  To break the assumptions of the local classifier, the
global one uses code analysis to identify \textit{init-only fields} and
\textit{fixed-length array type} according to the following definitions.

\begin{description}[leftmargin=0cm, listparindent=\parindent, itemsep=1pt, topsep=1pt]

\item[Init-only field.] A field of a non-primitive type \texttt{T} is
\textit{init-only}, if, for each object, this field will only be assigned 
once during the program execution. \footnote{We always treat the array element fields as non
init-only, otherwise the analysis needs to trace the element index value in each
assignment statement, which is not feasible in static code analysis.} 

\item[Fixed-length array types.] An array type \texttt{A} contained in the type-set
of field {\ttfamily \small f} is a \textit{fixed-length} array type w.r.t.
{\ttfamily \small f} if all the \texttt{A} objects assigned to {\ttfamily \small f}
are constructed with identical length values within a well-defined scope, such as a
single Spark job stage or a specific cached RDD. An example of symbolized constant
propagation is shown in Figure~\ref{code:cp}. Here, \texttt{array} is constructed
with the same length for whatever \texttt{foo()} returns. The fixed-length array types
with its element fields being SFST (or RFST) can be refined to SFST (or RFST).

\begin{figure}[!h]
\centering
\begin{lstlisting}
val a = input.readString().toInt() // a == Symbol(1)
val b = 2 + a - 1 // b == Symbol(1) + 1
val c = a + 1     // c == Symbol(1) + 1
if (foo()) array = new Array[Int](b)
else array = new Array[Int](c)
// array.length == Symbol(1) + 1
\end{lstlisting}
\caption{Symbolized constant propagation.}
\label{code:cp}
\end{figure}

\end{description}

In Figure~\ref{code:lr}, the {\ttfamily \small features} field is
only assigned in the constructor of \texttt{LabeledPoint} (lines 1--8), and
the length of {\ttfamily \small features.data} is a global constant value
{\ttfamily \small D} (lines 14-16). Thus, the size-class of
\texttt{LabeledPoint} can be refined to SFST.

\begin{algorithm}[!t]

\SetKwInOut{Input}{Input}\SetKwInOut{Output}{Output}
\SetKwData{SF}{\small{StaticFixed}}\SetKwData{RF}{\small{RuntimeFixed}}
\SetKwData{V}{\small{Variable}}
\SetKwProg{Fn}{Function}{}{end}
\SetKwFunction{SR}{SRefine}\SetKwFunction{RR}{RRefine}
\SetKwData{True}{\small{true}}\SetKwData{False}{\small{false}}

\Input{The top-level non-primitive type $T$; The locally-classified size-type $S_{local}$; Call graph of the current analysis scope $G_{call}$\;}
\Output{The refined size-type of $T$\;}
\lIf{\SR{$T, G_{call}$}}{\KwRet{\SF}}
\uElseIf{$S_{local}=$ \RF \bf{or} \RR{$T, G_{call}$}}{
  \KwRet{\RF}\;
}
\lElse{\KwRet{\V}}
\caption{Global classification analysis.}
\label{code:ga}
\end{algorithm}

Algorithm~\ref{code:ga} shows the procedure of the global classification.
The input of the algorithm is the target UDT and the call graph of the current
analysis scope. 
The refinement is done based on the following lemmas.

\newtheorem{lemma}{Lemma}
\begin{lemma}[SFST Refinement] An array type that is an RFST
or a VST can be refined to an SFST if and only if
for every array type in the type dependent graph, the followings
are true:
\begin{enumerate}[itemsep=1pt, topsep=1pt, parsep=1pt]
\item it is a fixed-length array type; and
\item every type in the type-set of its element field is an
SFST.
\end{enumerate}
\end{lemma}

\begin{lemma}[RFST Refinement] An array type that is a
	VST can be refined to an RFST if and only if:
\begin{enumerate}[itemsep=1pt, topsep=1pt, parsep=1pt]
\item every type in the type-sets of its fields is either an SFST or an RFST;
  and
\item each field with an RFST in its type-set is init-only.
\end{enumerate}
\end{lemma}

The call graph used for the analysis is built in the
pre-processing phase (Section~\ref{sec:impl}). 
The entry node of the call graph is the main method of the current analysis scope,
usually a Spark job stage, while all the reachable methods from the entry node as
well as their corresponding calling sequences are stored in the graph. 

\begin{algorithm}[!t]

\SetKwInOut{Input}{Input}\SetKwInOut{Output}{Output}
\SetKwData{SF}{\small{StaticFixed}}\SetKwData{RF}{\small{RuntimeFixed}}
\SetKwData{V}{\small{Variable}}\SetKwData{RD}{\small{RecurDef}}
\SetKwProg{Fn}{Function}{}{end}
\SetKwFunction{SR}{SRefine}
\SetKwData{FL}{\small{Fixed-Length}}
\SetKwData{True}{\small{true}}\SetKwData{False}{\small{false}}

\Fn{\SR{$t_{arg}, g_{arg}$}}{
  \Input{A non-primitive type $t_{arg}$; A call graph $g_{arg}$\;}
  \Output{\True or \False that $t_{arg}$'s size-type can be refined to \SF\;}
  \ForEach{field $f$ of type $t_{arg}$}{
    \ForEach{runtime type $t$ in $f.getTypeSet$}{
      \lIf{$t$ is not a primitive type {\bf and} not \SR{$t, g_{arg}$}}{\KwRet{\False}}
    }
  }
  \lIf{$t_{arg}$ is an array type {\bf and} $t_{arg}$ is not \FL in call graph $g_{arg}$}{
    \KwRet{\False}
  }
  \lElse{\KwRet{\True}}
}
\caption{Static fixed-sized type refinement: {\ttfamily SRefine($t_{arg}, g_{arg}$)}}
\label{code:sr}
\end{algorithm}

\begin{algorithm}[!t]

\SetKwInOut{Input}{Input}\SetKwInOut{Output}{Output}
\SetKwData{SF}{\small{StaticFixed}}\SetKwData{RF}{\small{RuntimeFixed}}
\SetKwData{V}{\small{Variable}}\SetKwData{RD}{\small{RecurDef}}
\SetKwProg{Fn}{Function}{}{end}
\SetKwFunction{SR}{SRefine}
\SetKwFunction{RR}{RRefine}
\SetKwData{IO}{\small{Init-Only}}\SetKwData{F}{\small{final}}
\SetKwData{True}{\small{true}}\SetKwData{False}{\small{false}}

\Fn{\RR{$t_{arg}, g_{arg}$}}{
  \Input{A non-primitive type $t_{arg}$; A call graph $g_{arg}$\;}
  \Output{\True or \False that $t_{arg}$'s size-type can be refined to \RF\;}
  \ForEach{field $f$ of type $t_{arg}$}{
    $analyze\_field \leftarrow$ \False\;
    \ForEach{runtime type $t$ in $f.getTypeSet$}{
      \If{$t$ is not a primitive type {\bf and} not \SR{$t, g_{arg}$}}{
        \uIf{\RR{$t, g_{arg}$}}{
          $analyze\_field \leftarrow$ \True\;
        }
        \lElse{\KwRet{\False}}
      }
    }
    \lIf{$analyze\_field$ {\bf and} $f$ is not \IO in call graph $g_{arg}$}{
      \KwRet{\False}
    }
  }
  \KwRet{\True}\;
}
\caption{Runtime fixed-sized type refinement: {\ttfamily RRefine($t_{arg}, g_{arg}$)}}
\label{code:rr}
\end{algorithm}

In line 7 of Algorithm~\ref{code:sr}, we use the following steps to identify the
fixed-length array types. (1) Perform the copy/constant propagation in the call
graph. The values passed from the outside of the call graph or returned by the I/O
operations will be represented by symbols considered as constant values. (2) For
a field {\ttfamily \small f} and an array type \texttt{A}, find all the
allocation sites of the \texttt{A} objects that are assigned to {\ttfamily
\small f} (i.e. the methods where these objects are created).  If all the length
values used in all these allocation sites are equivalent, \texttt{A} is of
fixed-length w.r.t. {\ttfamily \small f}.

In line 11 of Algorithm~\ref{code:rr}, we use the following rules to identify 
init-only or non-init-only fields: 1) a final field is init-only; 2) an array
element field is not init-only; 3) in addition, a field is init-only if it will
not be assigned in any method in the call graph other than the constructors of
its containing type, and it will only be assigned once in any constructor calling
sequence.

\subsection{Phased Refinement}

In a typical data parallel programming framework, such as Spark, each job can be
divided into one or more execution \textbf{phases}, each consisting of three
steps: (1) reading data from materialized (on-disk or in-memory) data
collectors, such as cached RDD, (2) applying an UDF on each data object, and (3) emitting the
resulting data into a new materialized data collector. 
Figure~\ref{code:phases-template} shows the framework of a job in
Spark. It consists one or more top-level computation loops, each reads data
object from its source, and writes the results into the sink.  Every
two successive loops are bridged by a data collector, such as an
RDD or a shuffle buffer.

We observe that the data-sizes of object types may have different levels of
variability at different phases. For example, in an early phase, data would be
grouped together by their keys and their values would be concatenated into an
array whose type is a VST at this phase. However, once the resulting objects are
emitted to a data collector, e.g. a cached RDD, the subsequent phases might not
reassign the array fields of these objects. Therefore, the array types can be
considered as RFSTs in the subsequent phases. We exploit this phenomenon to
refine a data type's size-class in each particular phase of a job, which is
called \emph{phased refinement}. This can be achieved by running the
global classification algorithm for the VSTs on each phase of the job.    

\begin{figure}[!t]
\centering
\begin{lstlisting}
// The first loop is the input loop.
var source = stage.getInput()
var sink = stage.nextCollection()
while (source.hasNext()) {
  val dataIn = source.next()
  ...
  val dataOut = ...
  sink.write(dataOut)  
}
// Optional inner loops
source = sink
sink = stage.nextCollection()
while (source.hasNext()) {...}
...
// The last loop is the output loop
source = sink
sink = stage.getOutput()
while (source.hasNext()) {...}
\end{lstlisting}
\caption{A code template of the Spark job stage.}
\label{code:phases-template}
\end{figure}
\section{Lifetime-based Memory Management}
\label{sec:lifetime} 

\subsection{The Spark Programming Framework}
\label{subsec:spark}

Spark provides a functional programming API, through which users can process Resilient
Distributed Datasets\linebreak (RDDs), the logical data collections partitioned across a cluster.
An important feature is that RDDs can be explicitly cached in the memory
to avoid re-computation or disk I/O overhead.

While Spark supports many operators, the ones most relevant for memory
management are some \textit{key-based} operators, including {\ttfamily \small
reduceByKey},
{\ttfamily \small groupByKey}, {\ttfamily \small join} and {\ttfamily \small sortByKey}
(analogues of GroupBy-Aggregation, GroupBy, Inner-Join and OrderBy in SQL).
These operators process data in the form of \textit{Key-Value} pairs.
For example, {\ttfamily \small reduceByKey} and {\ttfamily \small groupByKey}
are used for: 1) aggregating all \textit{Values} with the same \textit{Key} into
a single \textit{Value}; 2) building a complete \textit{Value} list for each
\textit{Key} for further processing. 

Furthermore, these operators are implemented using data shuffling. 
The shuffle buffer stores the combined value of each \text{Key}. For example,
for the case of {\ttfamily \small reduceByKey}, it stores a partial aggregate
value for each \text{Key}, and for the case of {\ttfamily \small groupByKey}, it
stores a partial list of \textit{Value} objects for each \text{Key}. When a new
\textit{Key-Value} pair is put into the shuffle buffer, eager combining is
performed to merge the new \textit{Value} with the combined value. 

For each Spark application, a driver program negotiates with the cluster resource
manager (e.g. YARN~\cite{vavilapalli:yarn}), which launches executors (each with
fixed amount of CPU and memory resource) on worker machines. An application can
submit multiple jobs. Each job has several \textit{stages} separated by data shuffles
and each stage consists of a set of tasks that perform the same computation. Each
executor occupies a JVM process and executes the allocated tasks concurrently in
a number of threads.

\subsection{Lifetimes of Data Containers in Spark}
\label{subsec:container}

In Spark, all objects are allocated in the running executors' JVM heaps, and their
references are stored in three kinds of \textit{data containers} described below.
A key challenge for Deca is deciding when and how to reclaim the allocated
space. In the \textit{lifetime} analysis, we focus on the end points of the lifetime of
the object references. The lifetime of an object ends once all its references
are dead.

\begin{description}[leftmargin=0cm, listparindent=\parindent, itemsep=2.3pt, topsep=2pt, parsep=1pt]

\item[UDF variables.] Each task creates function objects according
to its task descriptor. UDF variables include objects assigned to the fields of
the function objects and the local variables of their methods.
The lifetimes of the function object end when the running tasks
complete. 
In addition, as long-living objects are recommended to be stored in cached RDDs,
in most applications, local variables are dead after each method invocation.
Therefore, we treat all the data objects referenced only by the local variables
as short-living temporal objects.

\item[Cache blocks.] In Spark,
each RDD has an object that records its data source and the
computation function.
Only the cached RDDs will be materialized and retained in memory. A cached RDD
consists of a number of cache blocks, each being an array of objects. The
lifetimes of cached RDDs are explicitly determined by the invocations of
{\ttfamily \small cache()} and {\ttfamily \small unpersist()} in the
applications.  Whenever a cached RDD has been ``unpersisted'', all of its cache
blocks will be released immediately. For non-cached RDDs, the objects only
appear as local variables of the corresponding computation functions and hence
are also short-living.

\item[Shuffle buffers.] 
  A shuffle buffer is accessed by two successive phases in a job: one
  creates the shuffle buffer and puts data objects into it, while the other
  reads out the data for further processing. Once the second phase is completed,
  the shuffle buffer will be released.

With regard to the lifetimes of the object references stored in a 
shuffle buffer, there are three situations.  (1) In a sort-based shuffle buffer,
objects are stored in an in-place sorting buffer sorted by the \text{Key}.  Once
object references are put into the buffer, they will not be removed by the
subsequent sorting operations. Therefore, their lifetimes end when the shuffle
buffer is released.  (2) In a hash-based shuffle buffer with a {\ttfamily \small
reduceByKey} operator, the \textit{Key-Value} pairs are stored in an open hash
table with the \textit{Key} object as the hash key. Each aggregate operation
will create a new \textit{Value} object while keeping the \textit{Key} objects
intact. Therefore a \textit{Value} object reference dies upon an aggregate
operation over its corresponding \textit{Key}. (3) In a hash-based shuffle
buffer with a {\ttfamily \small groupByKey} operator, a hash table stores a set
of \textit{Key} objects and an array of \textit{Value} objects for each
\textit{Key}. The combining function will only append \textit{Value} objects to
the corresponding array and will not remove any object reference. Hence, the
references will die at the same time as the shuffle buffer. Note that these
situations cover all the key-based operators in Spark. For example, {\ttfamily
\small aggregateByKey} and {\ttfamily \small join} are similar to  {\ttfamily
\small reduceByKey} and {\ttfamily \small groupByKey} respectively. Other
key-based operators are just extensions of the above basic operators and hence
can be handled accordingly.

\end{description}

\begin{figure}[!t]
\centering
\includegraphics[width=0.42\textwidth]{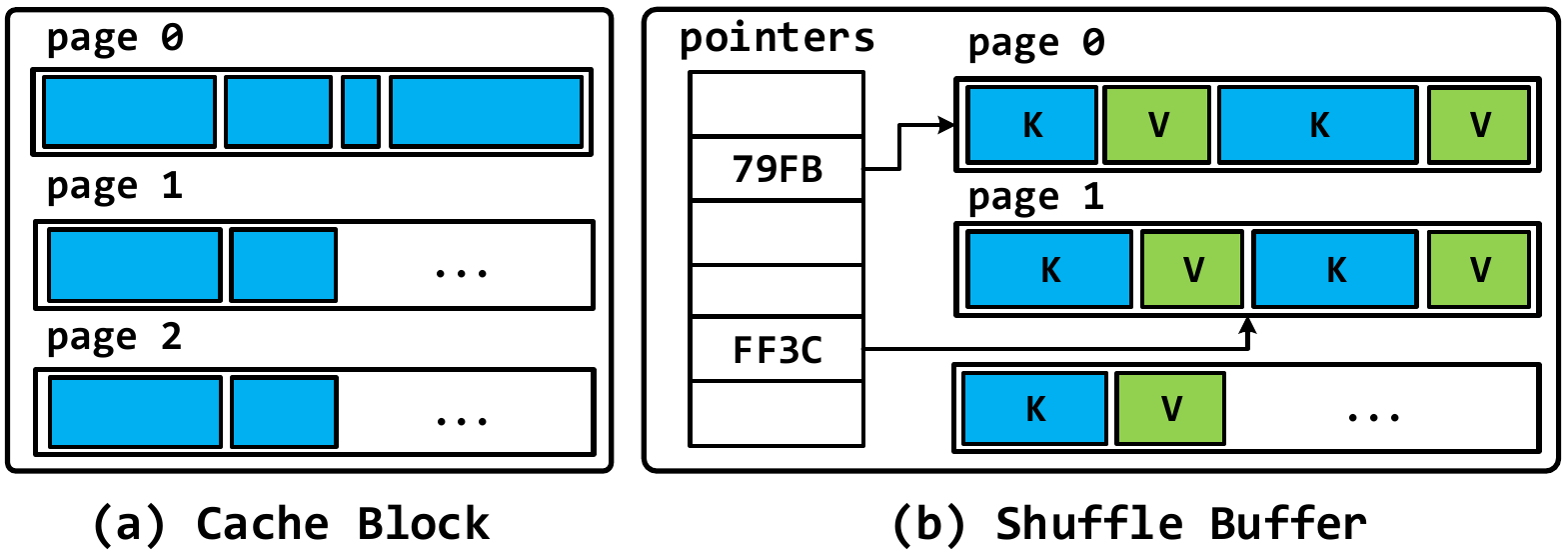}
\caption{Memory layouts of data containers.}
\label{fig:memory}
\end{figure}

\subsection{Data Containers in Deca}
As discussed above, object references'
lifetimes can be bound with the lifetimes of their containers.
Deca builds a data dependent graph for each job stage by points-to
analysis~\cite{lhot:spark} to produce the mapping relationships between all the
objects and their containers.  Objects are identified by either  their creation
statements if they are created in the current stage, or their source cached
blocks if they are read from cached blocks created by the previous stage.  

However, an object can be assigned to multiple data containers. For example, if
objects are copies between two different cached RDDs, then they can
be bound to the cached blocks of both RDDs. 
In such cases, we assign a sole \emph{primary container} as the owner of each
data object. Other containers are treated as \emph{secondary containers}. The
object ownership is determined based on the following rules:
\begin{enumerate}
\item Cached RDDs and shuffle buffers have higher priority of data ownership than UDF
variables, simply due to their longer expected lifetimes.
\item If there are objects assigned to multiple high-priority containers in the
  same job stage, the container created first in the stage execution will own these objects. 
\end{enumerate}

In the rest of this subsection, we present how data are organized within the
primary and secondary containers under various situations.  

\subsubsection{Memory Pages in Deca}
\label{subsec:memory}

Deca uses unified byte arrays with a common fixed size as logical memory pages
to store the decomposed data objects. A page can be logically split into
consecutive byte segments, one for each top-layer object. Each of such segment
can be further split into multiple segments, one for each lower-layer object,
and so on. The page size is chosen to ensure that there is only
a moderate number of pages in each executor's JVM heap so that the GC overhead
is negligible. On the other hand, the page size should not be too large either, so
that there would not be a significant unused space in the last page of a
container.

For each data container, a group of pages are allocated to store the objects it
owns. Deca uses a \textit{page-info} structure to maintain the metadata of each page group.
The page-info of each page graph contains: 1) \textit{pages}, a page array storing
the references of all the allocated pages of this page group; 2) \textit{endOffset}, an integer 
storing the start offset of the unused part of the last page in this group; 3)
\textit{curPage} and \textit{curOffset}, two integer values that store the progress
of sequentially scanning, or appending to, this page group. 

\subsubsection{Primary Container}

The way how Deca stores objects in their primary container depends on the
type of the container:

\begin{description}[leftmargin=0cm, listparindent=\parindent, itemsep=2pt, topsep=1pt, parsep=1pt]

\item[UDF variables.] Deca does not decompose objects owned by UDF variables.
These objects do not incur significant GC overheads, because: (1) the objects
only referenced by local variables are short-living objects and
they belong to the young generation, which will be reclaimed by the cheap minor GCs; (2) the
objects referenced by the function object fields may be promoted to the part of
old generation, but the total number of these objects in a task is relatively
small in comparing to the big input dataset.

\item[Cache blocks.] 
Deca always decomposes the SFST or RFST objects and stores their raw data bytes in the
page group of a cache block, while keeps the VST objects intact.
Figure~\ref{fig:memory}(a) shows the structure of a cache block of a cached RDD,
which contains decomposed objects.

A task can read objects from a decomposed cache block created in
a previous phase. If this task changes the data-sizes of these objects, Deca has to 
re-construct the objects and release the original page group. To avoid thrashing, when such
re-construction happens, Deca will not re-decompose these objects again even if
they can be safely decomposed in the subsequent phases.

\item[Shuffle buffers.] 
Figure~\ref{fig:memory}(b) shows the structure of a shuffle buffer.
Similar to cache blocks, data of an RFST or an SFST in a shuffle buffer will be
decomposed into the shuffle buffer's page group. However, unlike cached RDD, where data are
accessed in a sequential manner, data in a shuffle buffer will be randomly
accessed to perform sorting or hashing operations. Therefore, as illustrated on
the left-hand side of Figure~\ref{fig:memory}(b), we use an array to store the
pointers to the keys and values within a page. The hashing and sorting
operations are performed on the pointer arrays. However, the pointer array can
be avoided for a hash-based shuffle buffer with both the \textit{Key} and the
\textit{Value} being of primitive types or SFSTs.  This is because we can deduce
the offsets of the data within the page statically.

As we discussed in Section~\ref{subsec:container}, for a hash-based shuffle
buffer with a GroupBy-Aggregation computation, a combining operation would kill 
the old \textit{Value} object and create a new one. Therefore, \textit{Value}
objects are not long-living and frequent GC of these objects are
generally unavoidable. However, if the \textit{Value} object is of an SFST, then  
we can still decompose it and whenever a new object is generated by the
combining operation, we can just reuse the page segment occupied by the old
object, because the old and the new objects are of the same size. Doing this
would save the frequent GC caused by these temporary \textit{Value} objects. 

\end{description}

For brevity, we omit swapping data between memory and disks here. It is straightforward to adapt to the
cases with disk swapping for data caching and shuffling
(see Appendix~\ref{sec:swap} for details).

\subsubsection{Secondary Container}

There are common patterns of multiple data containers sharing the same data objects in
Spark programs, such as: 1) manipulating data objects in cache blocks or shuffle
buffers through UDF variables; 2) copying objects between cached RDDs; 3) immediately
caching the output objects of shuffling; 4) immediately shuffling the objects of
a cached RDD. 

If a secondary container is UDF variables, it will be assigned pointers to page
segments in the page group of the objects' primary container. Otherwise, Deca
stores data in the secondary container according to the following two different
scenarios:  (i) fully decomposable, where the objects can be safely decomposed
in all the containers, and (ii) partially decomposable, where the objects cannot
be decomposed in one or more containers.

\begin{figure}[!t]
\centering
\subfigure[Fully Decomp.]{
\label{fig:subfig:memory2}
\includegraphics[width=0.17\textwidth]{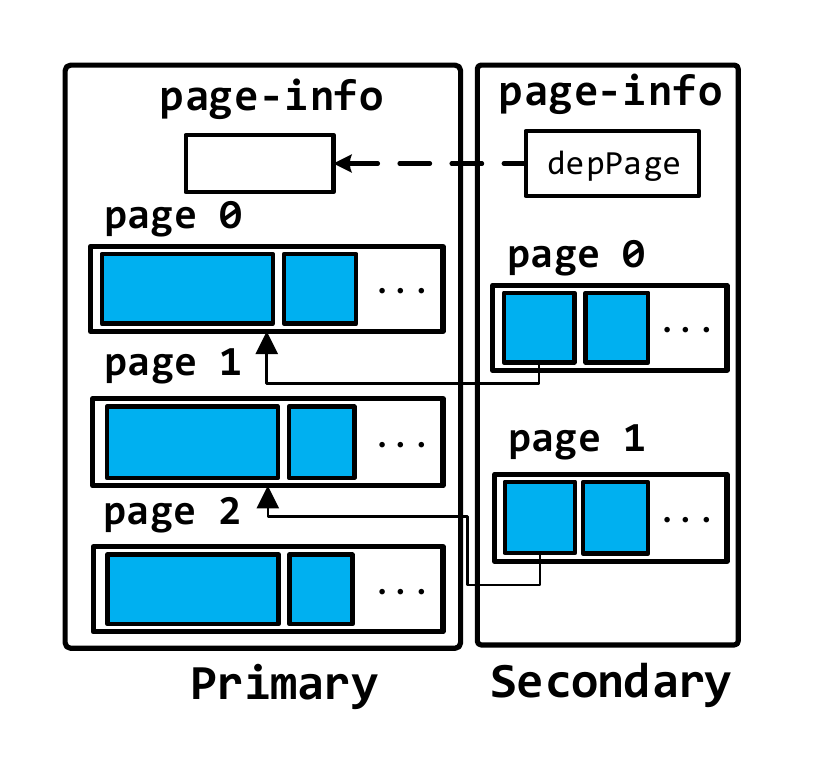}}
\subfigure[Partially Decomp.]{
\label{fig:subfig:layout-gbk}
\includegraphics[width=0.285\textwidth]{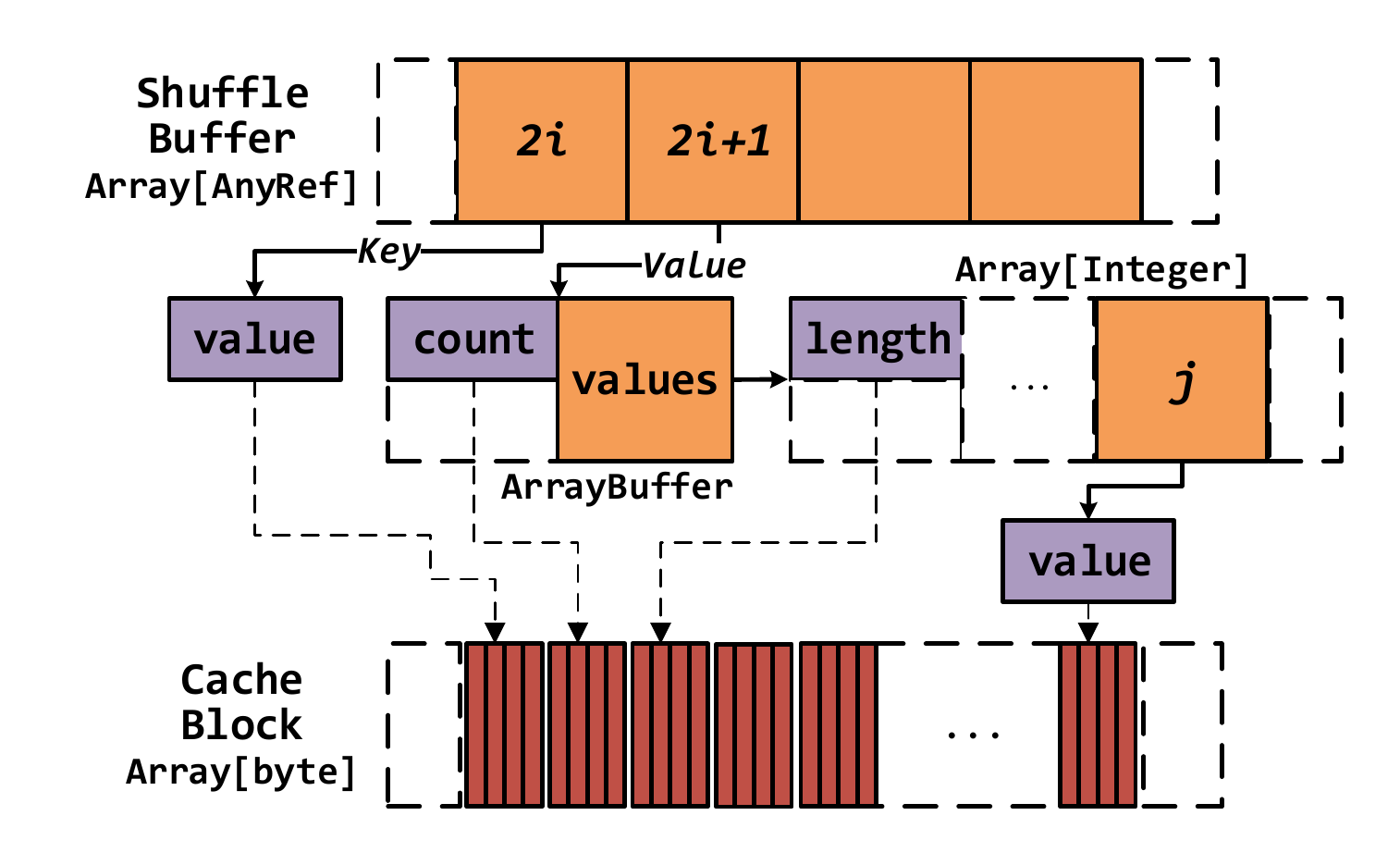}}
\vspace*{-4.4mm}
\caption{Examples of data layout in Deca.}
\vspace*{-4.5mm}
\label{fig:decomposable}
\end{figure}

\textbf{Fully decomposable.} This scenario is illustrated in
Figure~\ref{fig:subfig:memory2}. To avoid copy-by-value, a secondary container
only stores the pointers to the page group owned by the primary, one for each
object.  
Furthermore, we add an extra field, \textit{depPages}, to the page-info of the
secondary container to store the page-info(s) of the primary container(s). 

  Deca further performs optimizations for a special case, where a secondary
  container stores the same set of objects as the primary and does not
  require a specific data ordering. In such a case, Deca only generates a copy
  of the page-info of the page group owned by the primary container, and stores
  it in the secondary container. In this way, both containers actually share the
  same page group.  The memory manager uses a reference-counting method
  to reclaim memory space. Creating a new page-info of a page group increments
  its reference counter by one, while destroying a container (and its page-info)
  does the opposite. Once the reference counter becomes zero, the space of the
  page group can be reclaimed.

\textbf{Partially decomposable.} 
  In general, if the objects cannot be safely decomposed in one of the containers,
  then we cannot decompose them into a common page group shared by all
  the containers. However, if the objects are immutable or the modifications of
  objects in one container does not need to be propagated to the others,
  then we can decompose the objects in some containers and store the
  data in their object form in the non-decomposable containers. This is
  beneficial if the decomposable containers have long lifetimes.

Figure~\ref{fig:subfig:layout-gbk} depicts a representative example, where 
the output of a {\ttfamily \small groupByKey} operator, implemented via a
hash-based shuffle buffer, is immediately cached in a RDD.  Here, {\ttfamily
\small groupByKey} creates an array of \textit{Value} objects in the hash-based
shuffle buffer (see the middle of Figure~\ref{fig:subfig:layout-gbk}), and then
the output is copied to the cache blocks. The \textit{Value} array is of a VST
and hence cannot be decomposed in the shuffle buffer. However, in this case, the
shuffle buffers would die after the data are copied to the cache blocks, and the
subsequent modifications of the objects in the cache blocks do not need to be
propagated back to the shuffle buffers. Therefore, as shown in
Figure~\ref{fig:subfig:layout-gbk}, we can safely decompose the data in the
cache blocks, which have a long lifetime, and hence significantly reduce the GC
overhead.
\section{Implementation} \label{sec:impl}

We implement Deca based on Spark in roughly 6700 lines of Scala code. It consists
of an optimizer used in the driver, and a memory manager used in every
executor. The memory manager allocates and reclaims memory pages. It works together
with the Spark cache manager and shuffle manager, which manage the un-decomposed data
objects. The optimizer analyzes and transforms the code of each job when it is
submitted in the driver
(see Appendix~\ref{sec:hybrid} for details). 
The transformed code will use the API provided by the memory manager to create
pages and access the stored bytes. 

The Deca optimizer uses the Soot framework~\cite{www:soot} to analyze and
manipulate the Java bytecode. The Soot framework provides a rich set of
utilities, which implements classical program analysis and optimization methods.
The optimization consists of three phases: pre-processing, analysis and
transformation.

In the pre-processing phase, Deca uses \textit{iterator
fusion}~\cite{murray:steno} to bundle the iterative and isolated invocations of
UDFs into larger, hopefully optimizable code regions to avoid complex and costly
inter-procedural analysis. The per-stage call graphs and per-field type-sets are
also built using Soot in this phase. Building per-phase call graphs will be
delayed to the analysis phase if a phased refinement is necessary. In the
analysis phase, Deca uses methods described in Section~\ref{sec:classify} and
Section~\ref{sec:lifetime} to determine whether and how to decompose particular
data objects in their containers. Based on the obtained decomposability
information, new class files with transformed code will be generated and
distributed to all executors in the transformation phase. In general, the
accessing of primitive fields of decomposed data objects in the original code
will be transformed to accessing the corresponding page segments
(see Appendix~\ref{sec:trans} for details).
\section{Evaluation}
\label{sec:eval}

We use five nodes in the experiments, with one node as the master and the rest
as workers. Each node is equipped with two eight-core Xeon-2670 CPUs, 64GB
memory and one SAS disk, running RedHat Enterprise Linux 5 (kernel 2.6.18) and
JDK 1.7.0 (with default GC parameters). We compare the performance of Deca with
Spark 1.6. For serializing cached data in Spark, we use Kryo, which is a very
efficient serialization framework. 

Five typical benchmark applications in Spark are evaluated in these experiments:
\textit{WordCount} (WC), \textit{LogisticRegression} (LR), \textit{KMeans},
\textit{PageRank} (PR), \textit{ConnectedComponent} (CC).
As shown in Table~\ref{table:apps} they exhibit different characteristics and
hence can verify the system's performance in various different situations. 
For WC, we use the datasets produced by Hadoop \textit{RandomWriter} with
different unique key numbers (1M and 100M) and sizes (50GB, 100GB and 150GB). 
LR and KMeans use: 4096-dimension feature vectors (40GB and 80GB) extracted from
Amazon image dataset~\cite{mcauley:image}, and randomly generated 10-dimension
vectors (ranging from 40GB to 200GB).
For PR and CC, we use three real graphs: LiveJournal social
network~\cite{backstrom:livejournal} (2GB), webbase-2001~\cite{boldi:webgraph}
(30GB) and a 60GB graph generated by HiBench~\cite{www:hibench}. The maximum JVM
heap size of each executor is set to be 30GB for the applications with only data
caching or data shuffling, and 20GB for those with both caching and shuffling. 

\begin{table}[t]
\small
 \centering
    \begin{tabular}{ c || c | c | c | c}
    \hline
    \textbf{Application} & \textbf{Stages} & \textbf{Jobs} & \textbf{Cache} & \textbf{Shuffle} \\
    \hline
    WC & two & single & non & aggregated  \\
    \hline
    LR & single & multiple & static & non \\
    \hline
    KMeans & two & multiple & static & aggregated  \\
    \hline
    \tabincell{c}{PR \\CC} & multiple & multiple & static & \tabincell{c}{grouped \\aggregated} \\
    \hline
    
    \hline
    \end{tabular}
	\caption{Applications used in the experiments}
    \label{table:apps}
\end{table}

\subsection{Impact of Shuffling}

\begin{figure}[!t]
\centering
\subfigure[WC lifetime]{
\label{fig:subfig:jprofiler-wc}
\includegraphics[width=0.231\textwidth]{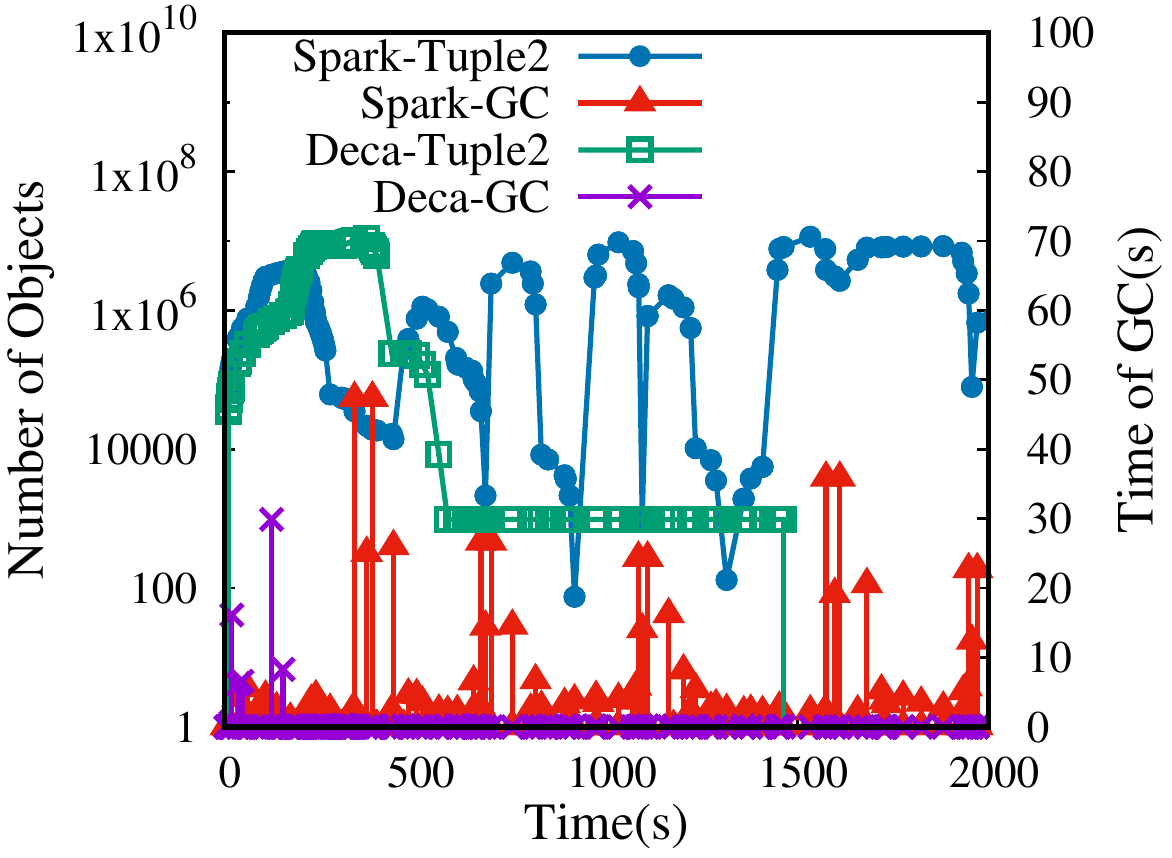}}
\hspace{-1.3ex}
\subfigure[WC exec]{
\label{fig:subfig:wc}
\includegraphics[width=0.231\textwidth]{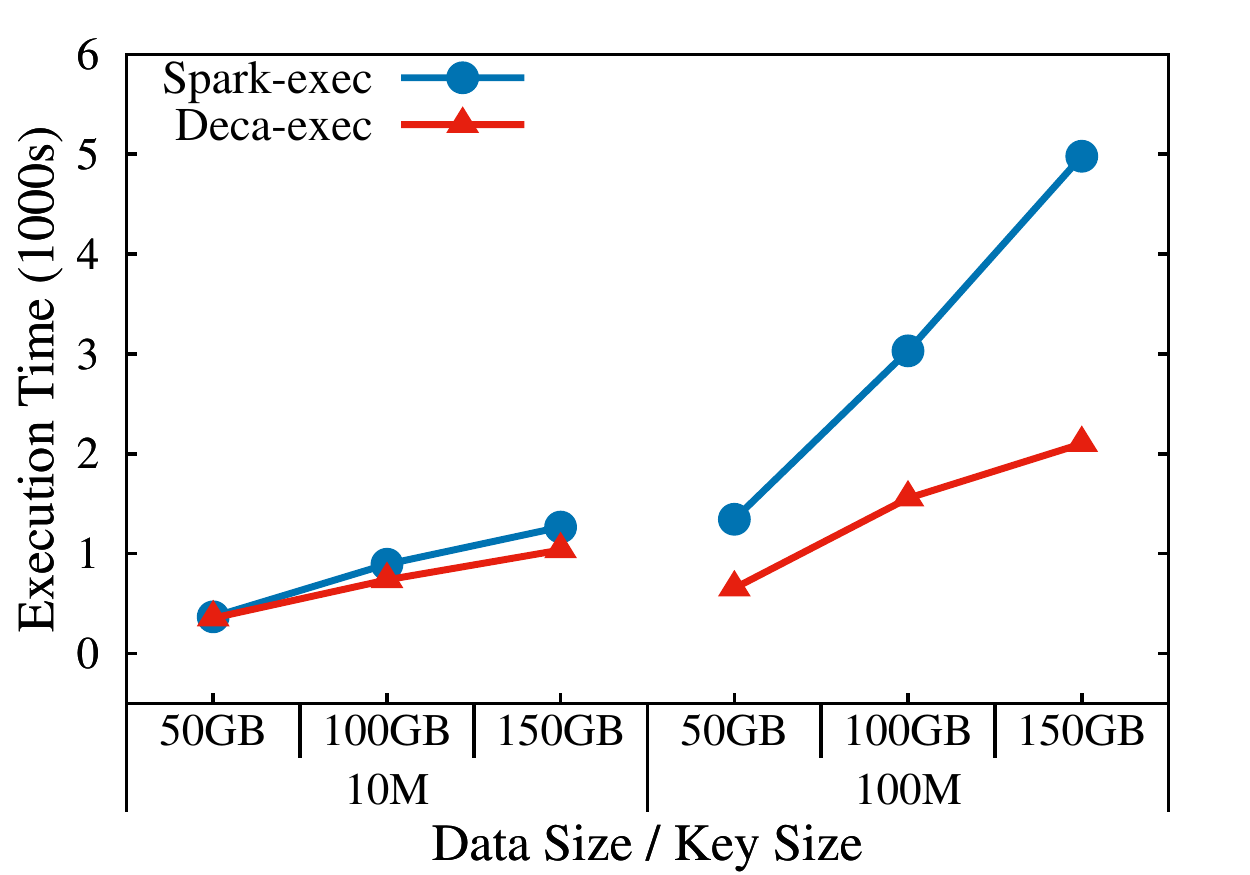}}
\label{fig:liftime}
\caption{Results of shuffling-only WC}
\end{figure}

WC is a two-stage MapReduce application with data shuffling between 
the ``map'' and ``reduce'' stages. 
We examine the lifetimes of data objects in the shuffle buffers with the
smallest dataset. We periodically record the alive number of objects and the GC
time with JProfiler 9.0. The result is shown in Figure~\ref{fig:subfig:jprofiler-wc}.
WC uses a hash-based shuffle buffer to perform eager aggregation, which is
implemented in \texttt{Tuple2}. The number of \texttt{Tuple2}
objects, which fluctuates during the execution, can indicate the number of
objects in shuffle buffers. While the number of \texttt{Tuple2} are also large in ''map'' stage but decrease in shuffle in Deca. 
GCs are triggered frequently to release the space
occupied by the temporary objects in the shuffle buffers.

\begin{figure*}[!t]
\centering
\subfigure[LR lifetime]{
\label{fig:subfig:jprofiler-lr}
\includegraphics[width=0.239\textwidth]{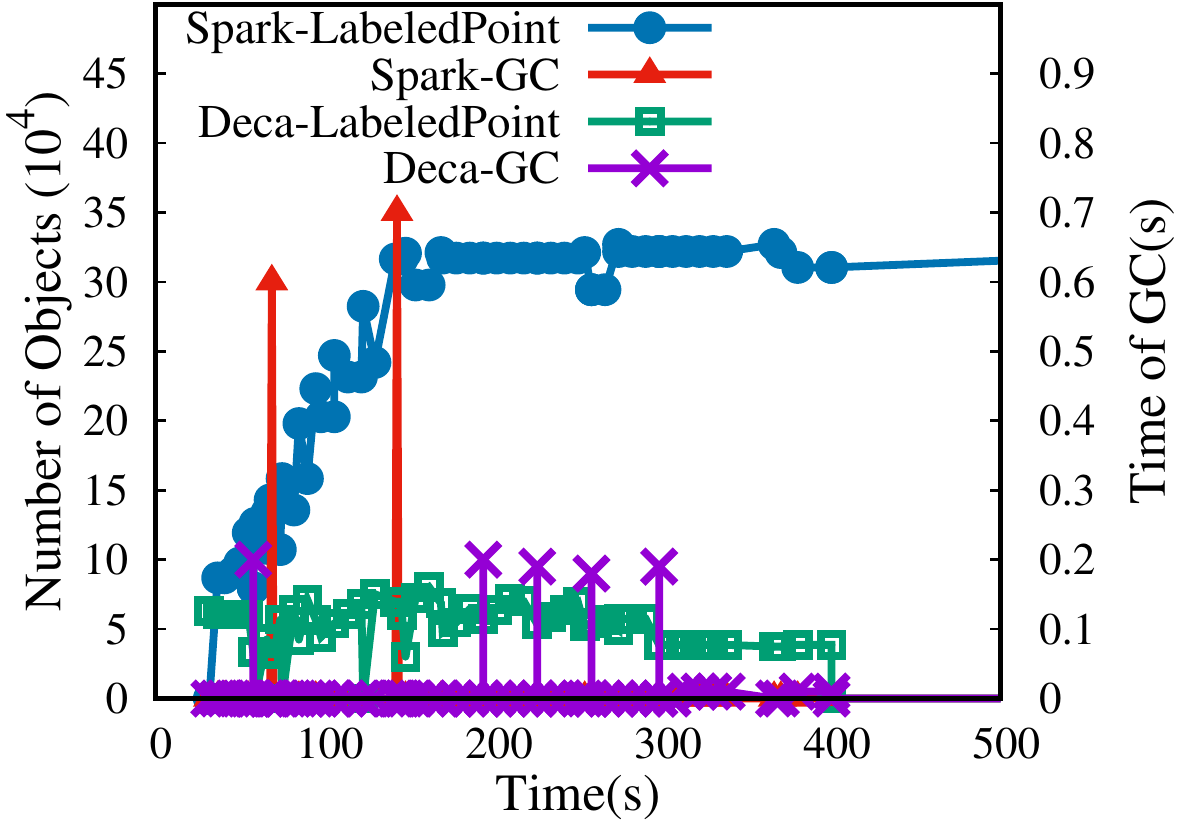}}
\subfigure[LR]{
\label{fig:subfig:lr}
\includegraphics[width=0.24\textwidth]{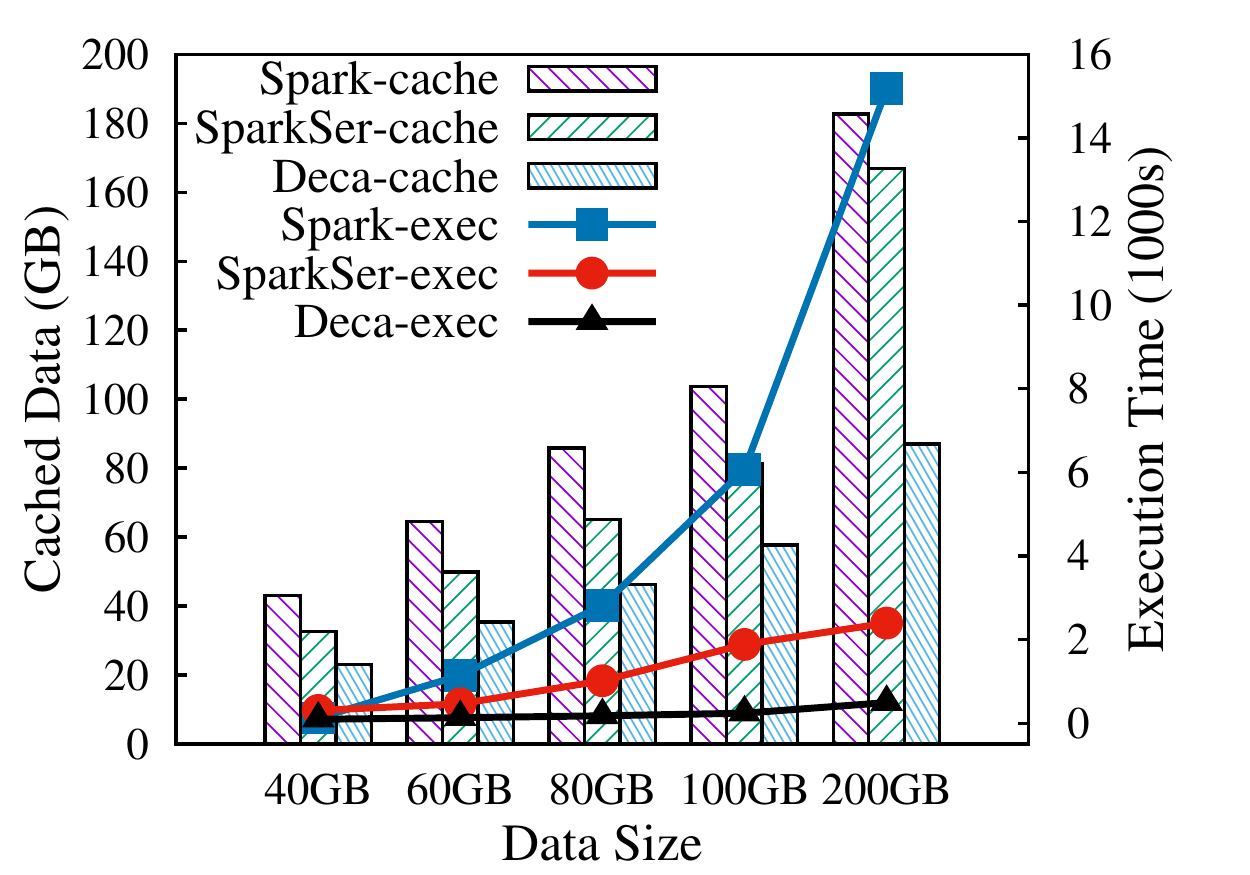}}
\subfigure[KMeans]{
\label{fig:subfig:kmeans}
\includegraphics[width=0.24\textwidth]{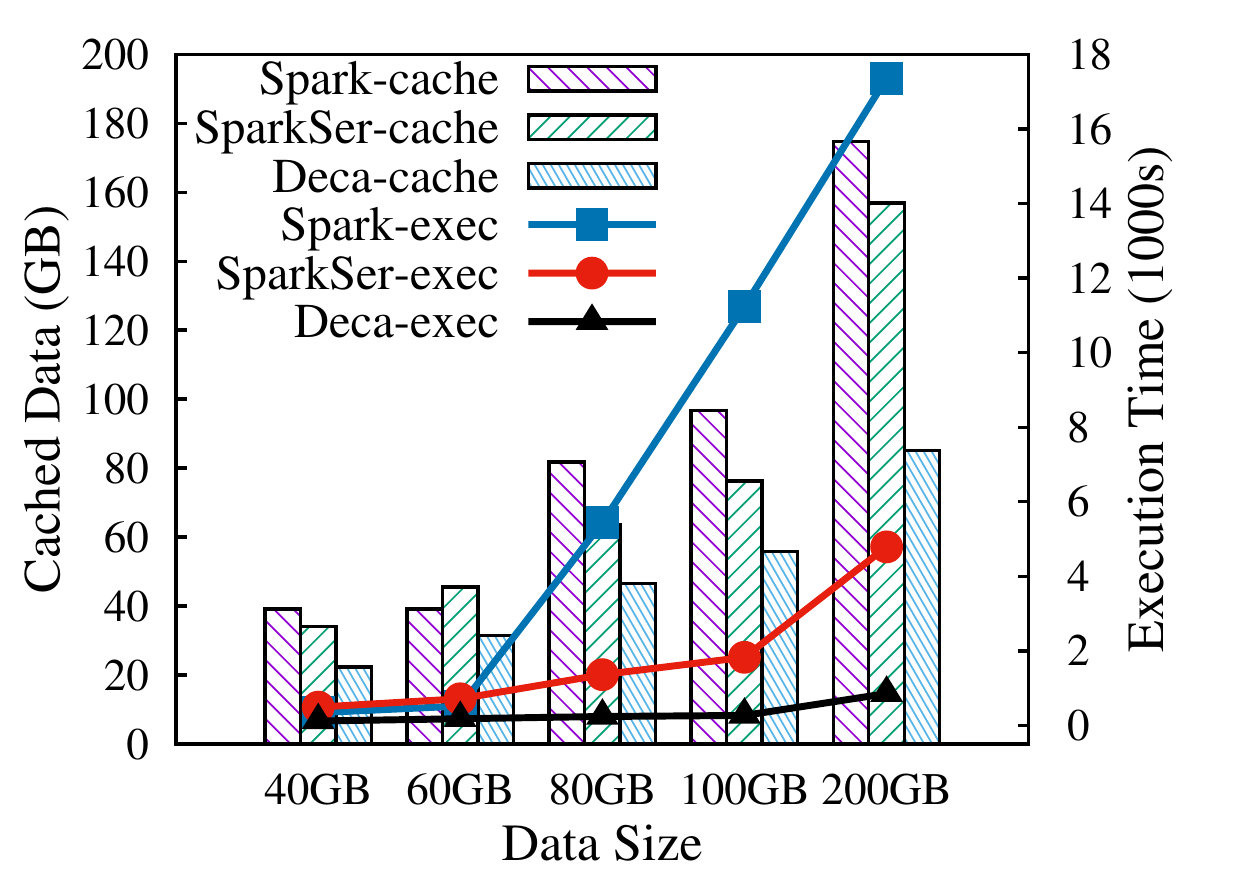}}
\subfigure[Amazon Image Dataset]{
\label{fig:subfig:lrandkmeans}
\includegraphics[width=0.238\textwidth]{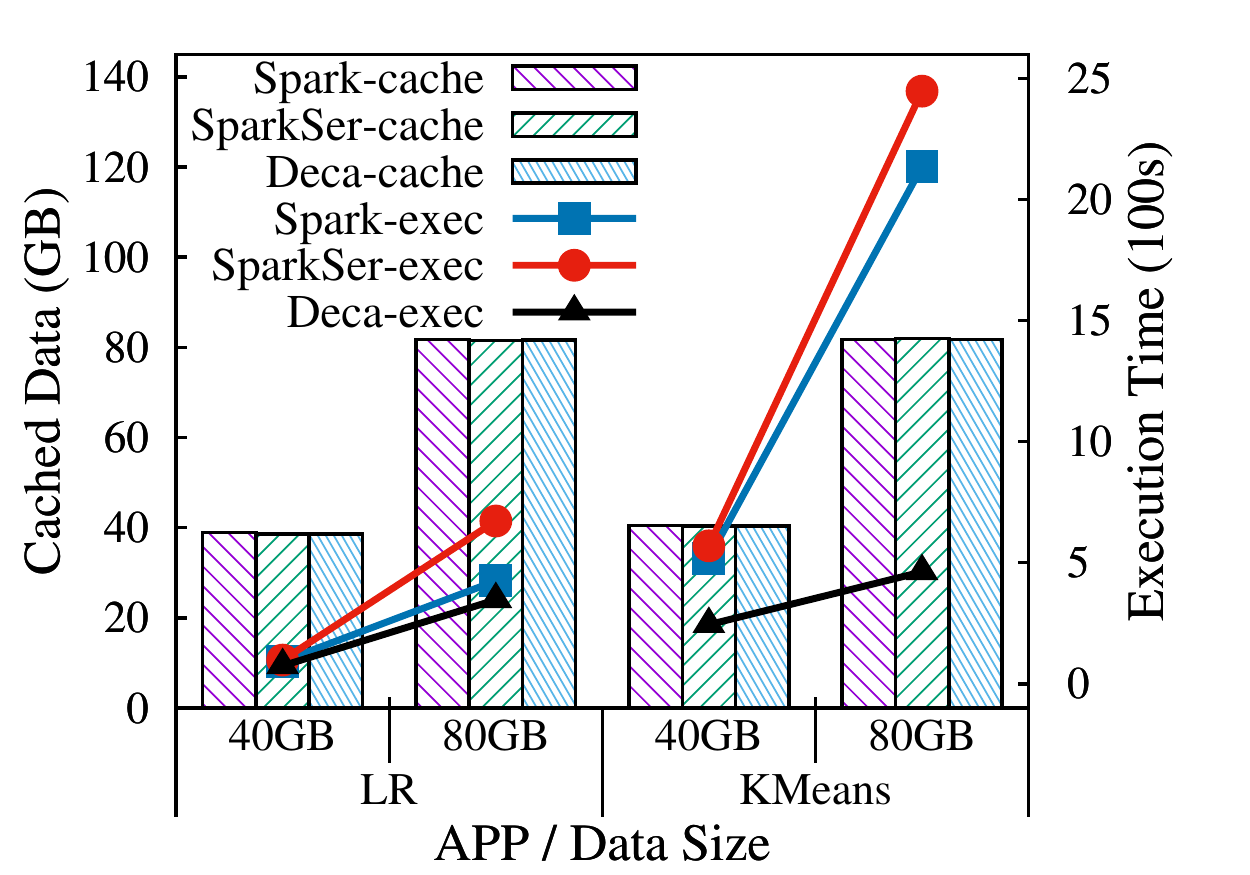}}
\caption{Results of caching-only LR/KMeans}
\label{fig:result1}
\end{figure*}

To avoid such frequent GC operations, Deca reuses the space occupied by the
partially-aggregated \textit{Value} for each \textit{Key} in the shuffle
buffer. Figure~\ref{fig:subfig:wc} compare the execution times of Deca and
Spark. In all cases, Deca can reduce the execution time by 10\%--58\%. One
could also see that the performance improvement increases with more number of
keys. This is because the size of a hash-based shuffle buffer with eager
aggregation mainly depends on the number of keys. The reduction of GC 
overhead would become more prominent with a larger number of keys. Furthermore,
since Deca stores the objects in the shuffle buffer as byte arrays, it also
saves the cost of data (de-)serialization by directly outputting the raw bytes. 

\subsection{Impact of Caching}

LR and KMeans are representative machine learning applications that perform
iterative computations. Both of them first load and cache the training dataset into
memory, then iteratively update the model until the pre-defined
convergence condition is met. In our experiments, we only run 30 iterations. We
do not account for the time to load the training dataset, because the iterative
computation dominates the execution time, especially consider that these
applications can run up to hundreds of iterations in a production environment.
We set $90\%$ of the available memory to be used for data caching.

We first examine the lifetimes of data objects in cache RDDs for
LogisticRegression (LR) using the 40GB dataset. The result is shown in
Figure~\ref{fig:subfig:jprofiler-lr}.  We find that the number of objects is rather
stable throughout the execution in Spark, but full GCs have been triggered several times
in vain (the peaks of the GC time curve). This is because most objects are
long-living and hence their space cannot be reclaimed. While these objects are less in Deca
because they are transformed to bytes after being read from the HDFS. Some objects still
live in old generation of JVM heap because no full gc is active.

By grouping massive objects with the same lifetime into a few byte arrays, Deca
can effectively eliminate the GC problem of repeatedly scanning alive data
objects for their liveness. Figure~\ref{fig:subfig:lr} and
Figure~\ref{fig:subfig:kmeans} show the execution times of LR and KMeans for
both Deca and Spark. Here we also examine the cases using Kryo to serialize the
cached data in Spark, which is denoted as ``SparkSer'' in the figures.

For the 40GB and 60GB datasets, the improvement is moderate and can be mainly
attributed to the elimination of object creation and minor GCs. In these cases,
the memory is sufficient to store the temporary objects, and hence full GC is
rarely triggered. Furthermore, serializing the cached data also helps reducing
the GC time. Therefore, with the 40GB dataset, SparkSer outperforms Spark by
reducing the GC overhead. However, for larger datasets, the overhead of data 
(de-)serialization cannot pay off the reduced GC overhead. Therefore, simply
serializing the cached data is not a robust solution.

For the three larger datasets the improvement is more significant. The speedups of
Deca are ranging from 16x to 41.6x. In these datasets, the long-living 
objects consume almost all available memory space, and therefore full GCs are
frequently triggered, which just repeatedly and unavailingly trace the cached
data objects in the old generation of the JVM heap. With the 100GB and 200GB
datasets, the additional disk I/O costs of cache swapping also prolong the
execution times of Spark. Deca keeps a smaller memory footprint of cached
data and swap smaller portion of data to the disks.

We also conduct the experiments on a real dataset, Amazon image dataset with
4096 dimensions.  Figure~\ref{fig:subfig:lrandkmeans} shows the speedups achieved by
Deca are ranging from 1.2x to 5.3x. With such a high dimensional dataset, the
size of object headers becomes negligible and therefore, the
sizes of the cached data of Spark and Deca are nearly identical. 

\subsection{Impact of Mixed Shuffling and Caching} 

\begin{table}[!h]
\scriptsize
  \centering
  	\begin{tabular}{| c | c | c | c | c |}
  	\hline
  	Graph & LiveJournal (LJ) & WebBase (WB) & HiBench (HB) \\
  	\hline
  	Vertices & 4.8M & 118M & 602M \\
  	\hline
  	Edges & 68M & 1B & 2B \\
  	\hline\hline
  	Data Size & 2GB & 30GB & 60GB \\
  	\hline
  	\end{tabular}
	\caption{Graph datasets used in PR and CC.}
    \label{table:graphdataset}
\end{table}

\begin{figure}[!t]
\centering
\subfigure[PR]{
\label{fig:subfig:pr}
\includegraphics[width=0.229\textwidth]{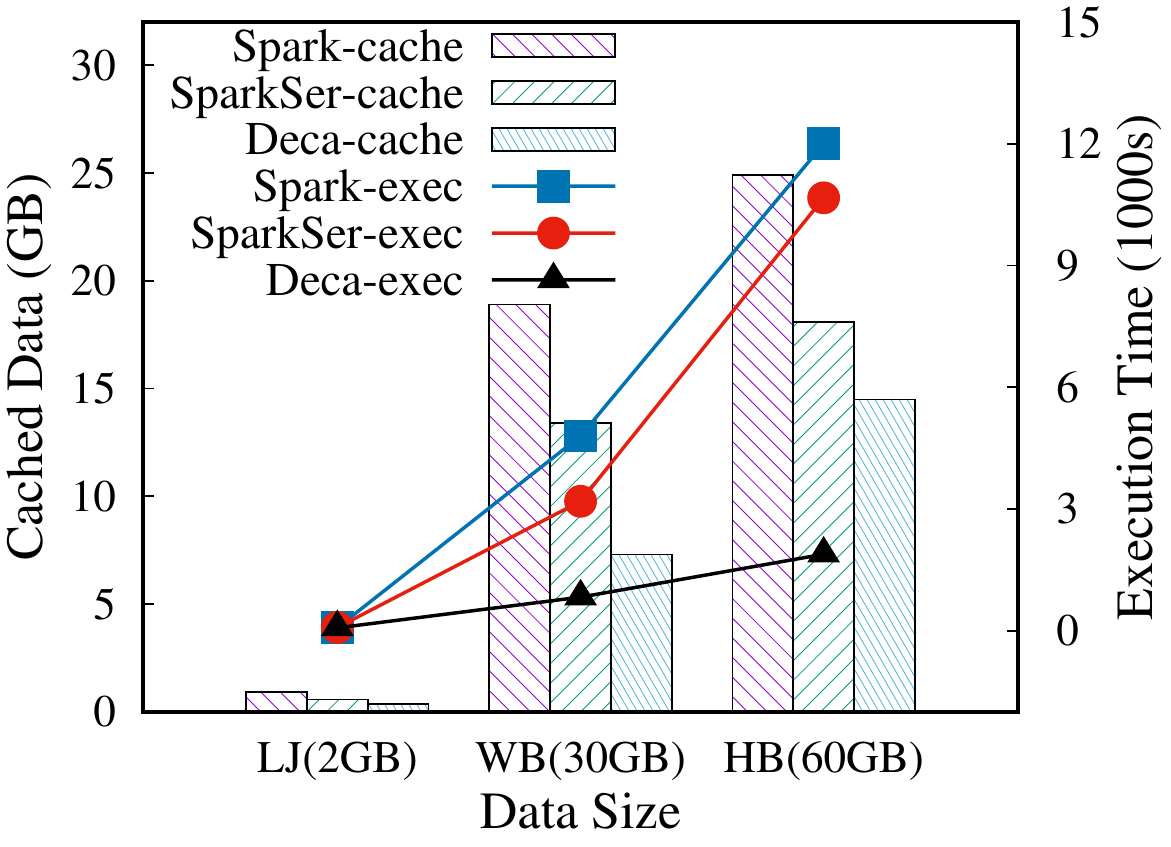}}
\subfigure[CC]{
\label{fig:subfig:cc}
\includegraphics[width=0.229\textwidth]{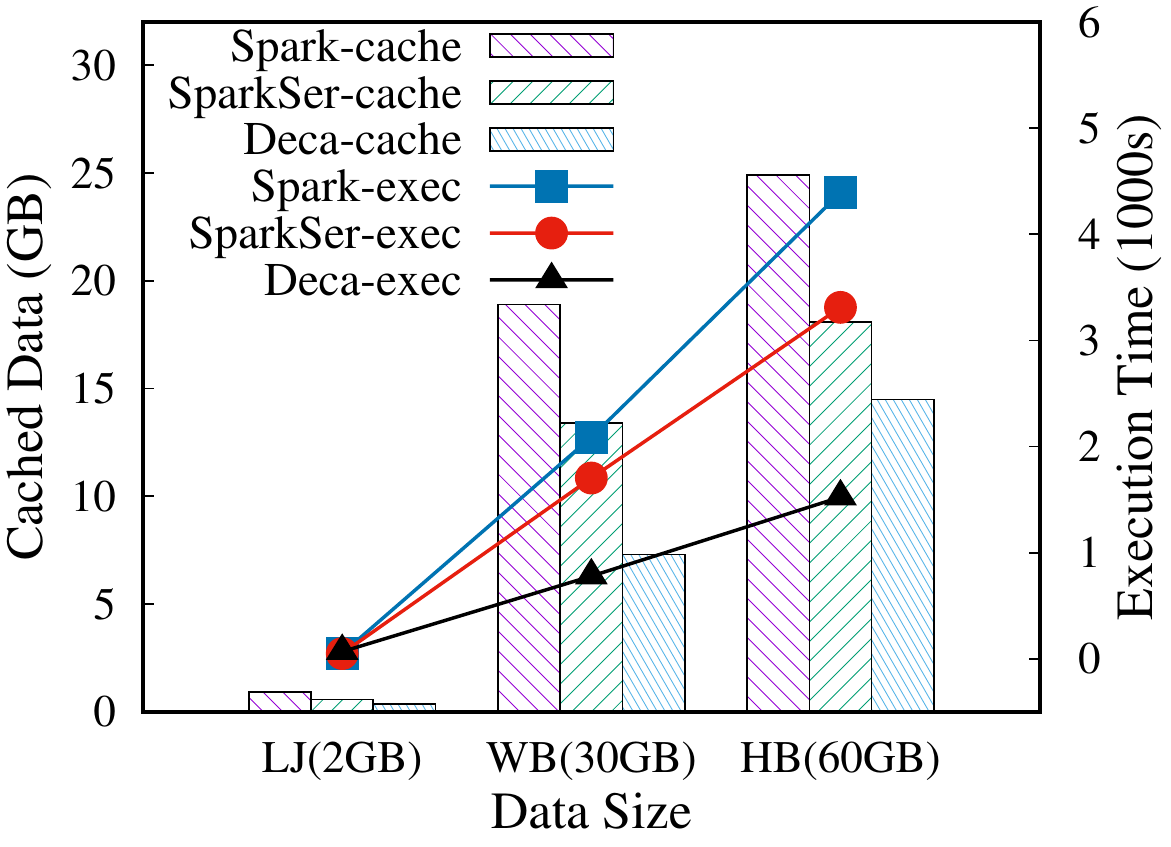}}
\caption{Results of PR and CC}
\label{fig:result2}
\end{figure}

PageRank (PR) and ConnectedComponent (CC) are representative iterative graph
computations. Both of them use {\ttfamily \small groupByKey} to transform the
edge list to the adjacency lists, and then cache the resulting data.  We use
three datasets with different edge numbers and vertex numbers as shown in
Table~\ref{table:graphdataset}. We set 40\% and 100\% of the available heap space
for caching and shuffling respectively. Edges will be cached during all
iterations, and shuffling is used in every iteration to aggregate messages for
each target vertex. We run 10 iterations in all the experiments. 

Figure~\ref{fig:subfig:pr} and Figure~\ref{fig:subfig:cc} show the execution times
of PR and CC for both Spark and Deca. The speedups of Deca are ranging
from 1.1x to 6.4x, which again can be attributed to the reduction of GC overhead and
shuffle serialization overhead. However, it is less dramatic than the previous
experiment. This is because each iteration of these applications creates new shuffle
buffers and releases the old ones. Then GC may be triggered to reclaim the memory
occupied by the shuffle buffers that are no longer in use. This reduces the memory
stress of Spark. We also see that SparkSer, which simply serialize the cache in
Spark, has little impact on the performance. The additional (de-)serialization
overhead offsets the reduction of GC overhead.

\subsection{GC improvement}

\begin{table}[!t]
  \scriptsize
  \centering
  \begin{tabular}{| c | c | c | c | c | c |}
  \hline
  \multirow{2}{*}{\textbf{App}} & \multicolumn{3}{| c |}{\textbf{Spark}} & \multicolumn{2}{| c |}{\textbf{Deca}} \\
  \cline{2-6}
   & \textbf{exec.} & \textbf{gc} & \textbf{ratio} & \textbf{gc} & \textbf{reduction} \\
  \hline
  
   WC: 150GB & 4980s & 2016s & 40.5\% & 12.2s & 99.4\% \\
   \hline
   LR: 80GB & 2820s & 2069.9s & 73.4\% & 2.5s & 99.9\% \\
   \hline  
   KMeans: 80GB & 5443s & 4294.8s & 78.9\% & 7.2s & 99.8\% \\
   \hline 
   PR: 30GB & 5544s & 3588.6s & 64.7\% & 21.7s & 99.4\% \\
   \hline 
   CC: 30GB & 2088s & 1443.9s & 69.2\% & 36s & 97.5\% \\
   \hline 
  
  \end{tabular}  
  \caption{GC time reduction.}
  \label{table:gccompare}

\end{table}

Table~\ref{table:gccompare} shows the times to run GC and the ratios of GC time
to the whole job execution time for the seven applications. For each
application, we only present the case with the largest input dataset that does
not have data swapping or spilling, to avoid the disk I/O affecting the
execution time. For each case, the GC time is an average of the values on all
executors. The result demonstrates the effect of GC elimination by Deca, and
how it improves the entire application performance. 

As shown in the result, the GC running time of LR and KMeans occupies the largest portion of the
total execution time among all cases, which are $73.4\%$ and $78.9\%$ respectively.
With the 80GB input dataset, the cached data objects almost consume all the
memory space of the old generation of the JVM heap. Deca reduces GC running
time in two ways: 1) smaller cache datasets trigger much less full GCs; 2) once a
full GC is triggered, the overhead of tracing objects is significantly
reduced.

Since all the other applications have shuffle phases in their executions, the
disk and network I/O account for a significant portion of the total execution
time.  Furthermore, reserving memory spaces for shuffle buffers makes that the
long-living cached objects occupy no more than $60\%$ of the total available
memory.  Therefore, in these cases the GC running time ranges from $64.7\%$ to
$69.2\%$.  This explains the different improvement ratios for different types of
applications reported above.

\begin{table}[!t]
  \scriptsize
  \centering
  \begin{tabular}{| c | c | c | c | c | c | c |}
  \hline
  \multirow{2}{*}{\textbf{App}} & \multicolumn{3}{| c |}{\textbf{Storage Fraction}} & \multicolumn{3}{| c |}{\textbf{GC algorithm}} \\
  \cline{2-7}
   & \textbf{frac.} & \textbf{exec.} & \textbf{gc} & \textbf{algo.} & \textbf{exec.} & \textbf{gc} \\
  \hline
  
   LR: 80GB & 0.8:0.2 & 2466s & 1918s & PS & 3102s & 2367s \\
   \cline{2-7}
   Deca:152s/ & 0.6:0.4 & 450s & 30s & CMS & 423s & 52s \\
   \cline{2-7}  
   1.6s & 0.4:0.6 & 606s & 19s & G1 & 332s & 22s \\
   \hline 
   PR:30GB & 0.4:1.0 & 5544s & 3588s & PS & 5544s & 3588s \\
   \cline{2-7} 
   Deca:828s/ & 0.1:1.0 & 3720s & 1532s & CMS & 6480s & 3506s \\
   \cline{2-7}  
   21.7s & 0.0:1.0 & 3804s & 1426s & G1 & 7440s & 1966s \\
   \hline 
  
  \end{tabular}
  \caption{GC tuning}
  \label{table:gctuning1}

\end{table}

We then compare Deca with GC tuning methods.
The Spark document~\cite{www:gc-tuning-spark} states that adjusting  
the fractions of memory allocated to cache blocks and to shuffle blocks is an
effective GC tuning method. Furthermore, we also compare with two GC algorithms
available in Hotspot JVM: namely CMS and G1. 
Table~\ref{table:gctuning1} shows the results. LR is very sensitive to GC
tuning. By setting fractions of cache  and shuffle buffer to $0.6$ and $0.4$ (the optimal based on
our experiments), respectively, or replacing PS with CMS or G1 with tuned
parameters, we can significantly improve the  performance of LR. However, PR is
much less sensitive to GC tunings, which is consistent with the previously reported
experiments~\cite{www:gc-tuning-databricks}.  However, we cannot achieve the
same performance gain by setting a higher number of concurrent GC threads in G1
as reported in~\cite{www:gc-tuning-databricks}.  We conjectures that it is
because of the difference of the configuration of the machines, which is not
stated in~\cite{www:gc-tuning-databricks}. This experiment
indicates that GC tuning is an effective way to improve GC performance in some
applications, however it is a cumbersome process and is highly dependent on the
applications and the system environment.

\begin{figure}[!t]
\centering
\subfigure[LR-40G]{
\label{fig:subfig:lr40g}
\includegraphics[width=0.15\textwidth]{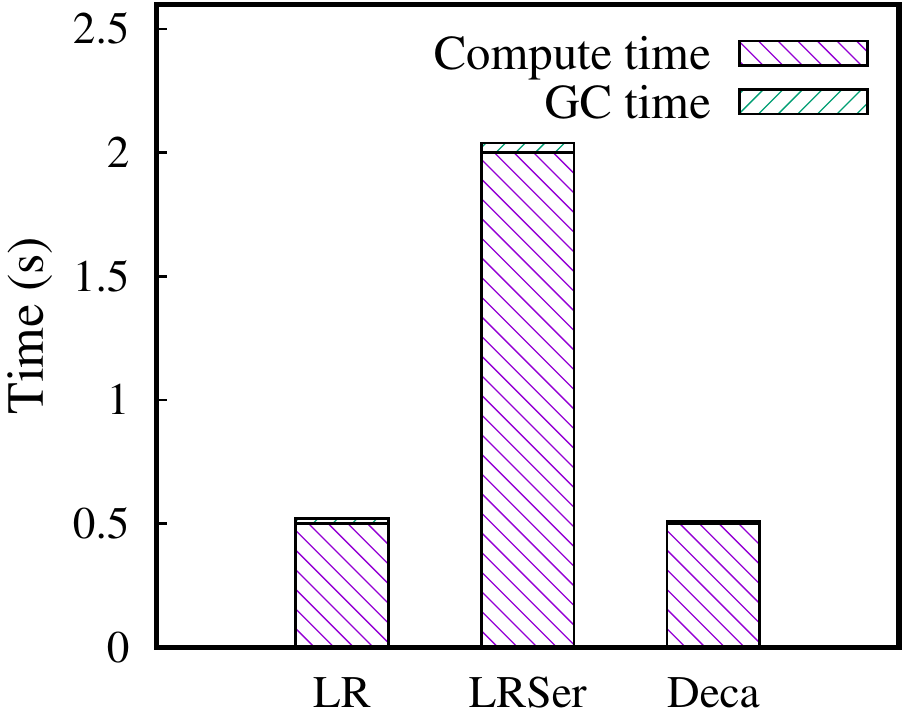}}
\hspace{-1.9ex}
\subfigure[LR-100G]{
\label{fig:subfig:lr100g}
\includegraphics[width=0.15\textwidth]{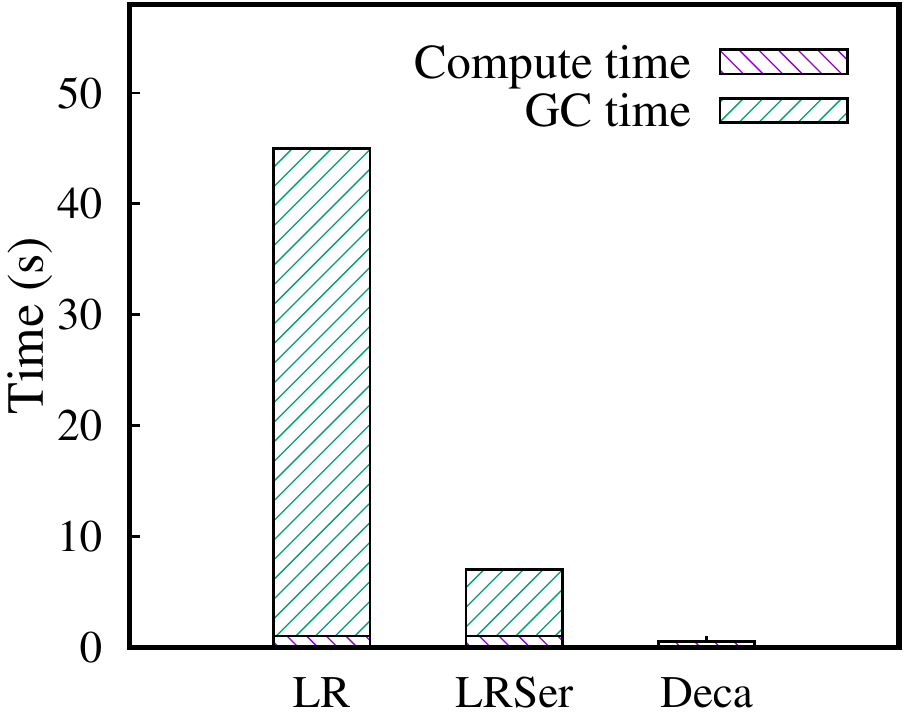}}
\hspace{-1.9ex}
\subfigure[PR-60G]{
\label{fig:subfig:pr60g}
\includegraphics[width=0.15\textwidth]{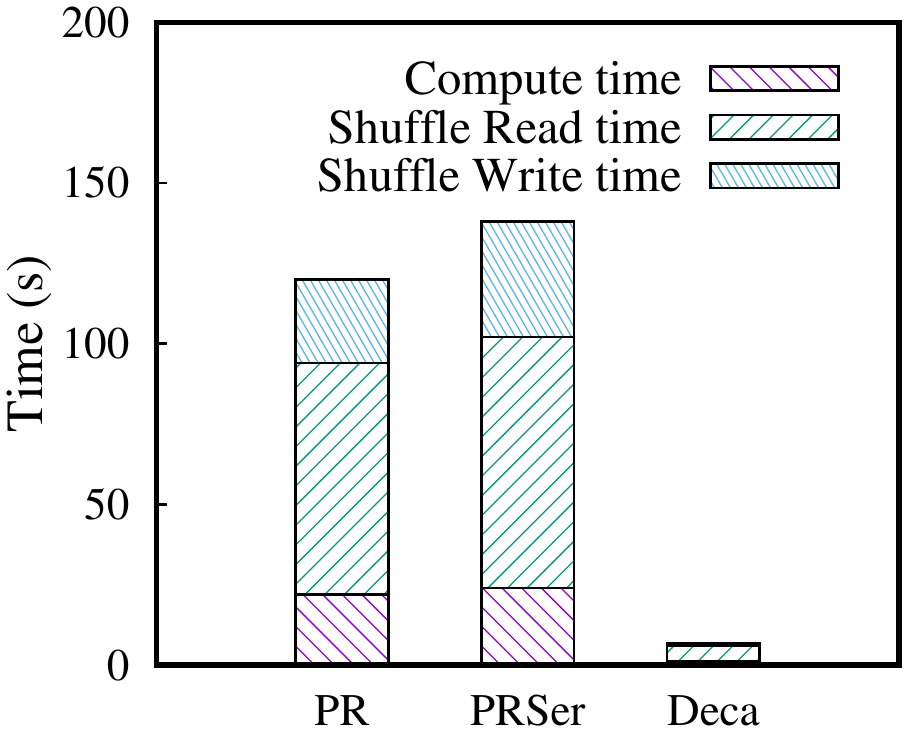}}
\caption{Breakdown of the task execution time.}
\label{fig:task-breakup}
\end{figure}

\begin{table}[!t]
  \scriptsize
  \centering
  \begin{tabular}{| c | c | c | c | c | c |}
  \hline
   \textbf{App} & \textbf{JVM heap} & \textbf{Time} & \textbf{Spark} & \textbf{Deca} & \textbf{SparkSer} \\
  \hline
  
   \multirow{4}{*}{\textbf{LR}} & \multirow{2}{*}{1.1GB} & exec. & 161.4s & 9.4s & 83.4s \\
   \cline{3-6}
    & & gc & 139.7s & 0.1s & 6.0s \\
   \cline{2-6}  
    & \multirow{2}{*}{20GB} & exec. & 10.9s & 9.5s & 44.6s \\
   \cline{3-6}
    & & gc & 0.02s & 0.18s & 0.04s \\
   \hline
   \multirow{4}{*}{\textbf{PR}} & \multirow{2}{*}{2GB} & exec. & 355.1s & 26.5s & 809.1s  \\
   \cline{3-6}
    & & gc & 100.8s & 0.18s & 505.1s \\
   \cline{2-6} 
    & \multirow{2}{*}{30GB} & exec. & 237.7s & 22.8s & 259.4s \\
   \cline{3-6}
    & & gc & 13.5s & 0.04s & 2.2s \\
   \hline
   \hline
   \multicolumn{3}{| c |}{Avg. time to serialize one object} & - & 3.7ms & 3.9ms \\
   \hline
   \multicolumn{3}{| c |}{Avg. time to de-serialize one object} & - & - & 27ms \\
   \hline 
  
  \end{tabular}
  \caption{Result of Microbenchmarks.}
  \label{table:microbenchmark}

\end{table}

\subsection{Microbenchmark}
To make a closer comparison, we attempt to break down the running
time of a single task in LR and PR in Figure~\ref{fig:task-breakup}. We use the
optimized memory fractions obtained in the previous subsection.  Note that tasks run
concurrently in the system and we present the slowest tasks in the
respective approaches, which somehow indicate the system bottleneck. 
In the LR-40G job, there is minimum GC overhead for all approaches, but the
deserialization overhead of SparkSer is obvious. For the LR-100G job, SparkSer
can also minimize GC overhead, but it needs to deserialize data into temporary
objects and hence still have some GC overhead. This shows the advantage of Deca'
code modification over serialization. Furthermore, for PR-60G, there is high
shuffling overhead in both Spark and SparkSer.  This is because the disk
swapping of the input cached RDD slows down the shuffle I/O.  Due to the smaller
footprint, Deca does not suffer from this problem. Note that GC and shuffle I/O
can run in parallel and shuffling is the bottleneck in this setup.

To further analyze the CPU overhead, we run LR and PR within a controlled
environment to eliminate the impact of memory footprint and other non-CPU
factors, such as the Spark's task scheduling delay and shuffling I/O. This is
done by using a multi-threaded Java program in a single machine to emulate the
workflow of Spark without task scheduling and shuffling I/O.  

In the LR job, we use 8 million randomly-generated 10-dimensional labeled
feature vectors as the input dataset. The input is
first partitioned and cached in object arrays (Spark) or byte arrays
(SparkSer and Deca). The cached partitions are then evenly dispatched to
computing threads. The number of iterations
is set to 50.  The results in Table~\ref{table:microbenchmark} show that, when
the heap is large enough (20 GB) and hence there is negligible GC overheads,
Deca is almost identical to Spark but SparkSer has a poor performance due to the
high deserialization overhead. Furthermore, when the JVM heap size is relatively
small (1.1 GB), Spark suffers from high GC overheads while both SparkSer and
Deca can keep the GC overheads low. Again, Deca outperforms SparkSer because it
does not require de-serialization and has a lower GC overhead. We also measure
the average time for Deca and Kryo to serialize and de-serialize each object.
The results are reported at the bottom of Table~\ref{table:microbenchmark}. We
can see that Deca has a similar serialization cost as Kryo, while Deca does not
involve a significant de-serialization overhead as Kryo does.   
 
A similar experiment is done on PR, which uses both cache and shuffle buffers.
The Pokec graph~\cite{www:snap} with 1.6M vertices and 30M edges is used as the
input. Note that Spark does not support in-memory serialization for shuffle
buffers, so SparkSer only serializes the cached data.  As shown in
Table~\ref{table:microbenchmark}, when the GC overhead is negligible with a
large heap, SparkSer again suffers from high de-serialization overhead, while
Deca runs significantly faster than Spark. This is because Spark needs to access
auto-boxed objects in generic-type containers in the shuffle buffers, while Deca
directly operates on the primitive values (note that this is not the case for
LR, because LR does not involve shuffling.).  When the GC overhead is high with
a smaller heap, SparkSer works worse than Spark because of the {\ttfamily \small
join} operation in PR. During the {\ttfamily \small join} operation, SparkSer
de-serializes the cached data and stores the resulting objects in the shuffle
buffers. On the other hand, Spark just stores references of the cached objects
in the shuffle buffers. Thus, SparkSer suffers from even higher GC overheads
than Spark does. Here, Deca's superiority is attributed to its ability to
decompose not only cache blocks but also shuffle buffers.

In summary, Deca has lower GC overhead, smaller footprint, no data deserialization,
and no data boxing and un-boxing. All these factors can be important to job
execution time, and their significance depends on the actual scenarios.
  
\subsection{Comparing with Spark SQL}
 
In this experiment, we compare Deca with Spark SQL, which is optimized for
SQL-like queries. Spark SQL uses a serialized column-oriented format to store
in-memory tables, and, with the project Tungsten, the shuffled data of certain
(built-in or user-defined) aggregation functions, such as AVG and SUM, are also
stored in memory with a serialized form. The serialization can be either
auto-generated or manually written by users. We use a dataset sampled from the
Common Crawl document corpus, where \textit{rankings} is 6.6GB and \textit{uservisits}
is 44GB. We use the table schemas and the two SQL queries provided in Spark's
benchmark~\cite{www.sparkbenchmark}. For each query, a
semantic-identical hand-written
Spark program (with RDDs) are used for Spark and Deca. The input tables are
entirely cached in memory before being queried. We disable the in-memory compression
of Spark SQL.

\lstset{language=SQL,
        frame=none,
        aboveskip=1.8mm,
        belowskip=1.8mm,
        xleftmargin=2em,
        basicstyle={\linespread{1}\fontsize{8}{9}\ttfamily},
        numbers=none}

The first query is a simple filtering: 

\begin{lstlisting}
SELECT pageURL, pageRank FROM rankings
    WHERE pageRank > 100;
\end{lstlisting}

The second one is a typical GroupBy aggregate query: 

\begin{lstlisting}
SELECT SUBSTR(sourceIP, 1, 5), SUM(adRevenue)
    FROM uservisits
    GROUP BY SUBSTR(sourceIP, 1, 5);
\end{lstlisting}

\begin{table}[!t]
 \small
  \centering
  \begin{tabular}{| c | c | c | c | c |}
  \hline
   & \textbf{App} & \textbf{exec.} & \textbf{gc} & \textbf{cache} \\
  \hline
  \multirow{3}{*}{\textbf{Query1}} & Spark & 23s & 2.23s & 18.6GB \\
  \cline{2-5}
   & Spark SQL & 22s & 0.49s & 5.6GB \\
  \cline{2-5}
   & Deca & 22s & 0.26s & 6.3GB \\
  \hline
  
  \multirow{3}{*}{\textbf{Query2}} & Spark & 396s & 192.4s & 97.9GB(23.1GB) \\
   \cline{2-5}
   & Spark SQL & 180s & 3.0s & 47.1GB \\
  \cline{2-5}
   & Deca & 192s & 4.2s & 55.6GB \\
  \hline
  
  \end{tabular}
  \caption{Execution times of two exploratory SQL query in Spark, Spark SQL and
  Deca. In Query2, the size of swapped cache data in Spark is 23.1GB.}
  \label{table:SparkSQL1}
\end{table}

The results are shown in
Table~\ref{table:SparkSQL1}.
All three systems perform equally well for the simple filtering query with small input
table. Although the GC running time in Spark is higher than that in the other two systems, it
only accounts for a negligible portion of the total execution time. For the second query
with a larger table, the GC overhead is significant for Spark. We can see that,
similar to Spark SQL, Deca can reduce more than 50\% of the execution time in
comparing to Spark, while keeping the generality of Spark's programming
framework.
\section{Related Work}
\label{sec:related}

The inefficiency of memory management in managed runtime platforms for big data
processing systems has been widely acknowledged. The existing efforts can be
categorized into the following directions.

\begin{description}[leftmargin=0cm, listparindent=\parindent, itemsep=1pt, topsep=1pt, parsep=1pt]

\item[GC tuning.] Most traditional GC tuning techniques are proposed for
  long-running latency sensitive web servers. Some open source distributed NoSQL
  systems, such as Cassandra~\cite{www:gc-tuning-cassandra1} and
  HBase~\cite{www:gc-tuning-hbase1}, use these latency-centric methods to avoid
  long GC pauses by replacing the Parallel GC with, e.g. CMS or G1 and tuning their parameters.  

\item[GC algorithms.] Implementing better GC algorithms is another line of
  work. Maas et al.~\cite{maas:trashday} propose a holistic runtime system for
  distributed data processing that coordinates the GC executions on all workers
  to improve the overall job performance. Gidra et al.~\cite{gidra:numagic}
  propose a NUMA-aware garbage collector
  for data processing running on machines with a large memory space. These
  approaches' requirements of modifying JVMs prevent them being adopted on production
  environments.
  On the other hand, Deca employs a non-intrusive approach and requires
  no JVM modification.

\item[Object serialization.] Many distributed data processing systems, such as Hadoop, Spark
  and Flink, support serializing in-memory data objects into byte arrays.
However, object serialization has long been acknowledged as having a high
overhead~\cite{carpenter:ser,welsh:ser,miller:pickle}.
Deca transforms the program code to directly access the raw data stored in byte
arrays and avoids such overheads.

\item[Region-based memory management (RBMM).] In\linebreak RBMM~\cite{tofte:region},
  all objects are grouped into a set of hierarchical regions, which
  are the basic units for space reclamation. The reclamation of a region
  automatically triggers the recursive reclamation of its sub-regions. This
  approach requires the developers to explicitly define the mapping from objects
  to regions, as well as the hierarchical structures of regions.
Gog et al.~\cite{gog:broom} report the early work on using RBMM for distributed
data processing that runs on .NET CLR. However, the evaluation is conducted
using task emulation with manually implemented operators, while the details
about how to transparently integrate RBMM with user codes remain unclear.
Nguyen et al.~\cite{nguyen:facade} propose an approach that stores all alive
objects of user-annotated types in a single region managed by a simplified version
of RBMM, thereby bypassing the garbage collection. The occupied space of data
objects will be reclaimed at once at a user-annotated reclamation point.
This method is unsuitable for systems that create data objects with diverse
lifetimes such as Spark. Deca can be regarded as a variant of RBMM, which
automatically maps objects with similar lifetimes to the same region. 

\item[Domain specific systems.] Some domain-specific data-\linebreak parallel
systems make use of its specific
computation structure to realize more complex memory management. Since the
early adoption of JVM in implementing SQL-based data-intensive
systems~\cite{shah2001java}, efforts have been devoted to making use of the 
well-defined semantics of SQL query operators to improve the memory
management performance in managed runtime platforms.
Spark SQL~\cite{armbrust:sparksql} transforms relational tables to serialized
bytes in a main-memory columnar storage. Tungsten~\cite{www:tungsten}, a Spark
sub-project, enables the serialization of hash-based shuffle
buffers for certain Spark SQL operators. 
Deca has a similar performance as Spark SQL for
structured data processing, meanwhile it provides more flexible computation and data
models, which eases the implementation of advanced iterative applications such
as machine learning and graph mining algorithms.

\end{description}
\section{Conclusion}

In this paper, we identify that GC overhead in distributed data processing systems
is unnecessarily high, especially with a large input dataset.
By presenting Deca's techniques of analyzing the variability of object sizes and
safely decomposing objects in different containers, we show that it is possible
to develop a general and efficient lifetime-based memory manager for distributed
data processing systems to largely eliminate the high GC overhead. The
experiment results show that Deca can significantly reduce Spark's application
running time for various cases without losing the generality of its programming
framework. Finally, to take advantage of Deca's optimization, a user is
recommended to not creating a massive number of long-living objects of a VST,
which cannot be safely decomposed.

\section{Acknowledgments}

We want to thank Beng Chin Ooi of the NUS,
Xipeng Shen of the NCSU, and Bingsheng He of the NTU for their valuable comments.

\balance

\bibliographystyle{abbrv}
{
  \bibliography{ref}}

\begin{appendix}

\section{Hybrid Optimization}
\label{sec:hybrid}

Intuitively, Deca can be implemented as a standalone tool that transforms the
compiled jar files of a Spark program before its execution. However, a 
Spark driver program may execute many jobs, each consisting of several stages
separated by shuffles. The job submission will be implicitly triggered by an
\emph{action}, such as \texttt{reduce}, which returns a value to the driver
after running a UDF on a dataset. According to the results returned by the
current job, the driver decides how to submit the next job.

A driver program can freely use the control statements (if/for/while)
to control the computation. Therefore, it may submit different jobs with
different input datasets and configuration parameters. A static optimization has
to enumerates and processes all the possible jobs by exhaustively exploring a
large number of possible execution paths of the program, which is the well-known 
\textit{path explosion} problem.
This is even infeasible if the program has loop structures, which render the
number of execution paths unbounded.

To address these challenges, we implement Deca in a hybrid way, 
which contains a static analyzer and a runtime optimizer.
The static analyzer extracts priori knowledge about the UDFs and UDTs of the
target programs, which can be used to reduce the runtime optimization overheads.
The runtime optimizer intercepts the submitted jobs at runtime, and optimizes
each job before actually submitting it to the Spark platform.  With this
approach, Deca only optimizes the actually submitted jobs, and thereby
completely eliminates the need for exploring all the execution paths.

\section{Code Transformation}
\label{sec:trans}

In each stage, for the data objects of the UDT that can be safely decomposed,
Deca transforms the corresponding code and leaves the unoptimizable part
unchanged. The transformation phase can be further split into three sub-phases,
which are described below.

\begin{description}[leftmargin=0cm, listparindent=\parindent, itemsep=1pt, topsep=4pt, parsep=1pt]

\item[Decomposition.] In this sub-phase, Deca generates a synthesized class for
  each UDT (called \textbf{SUDT}) to access the decomposed raw data. Logically,
  the {\ttfamily \small this} reference of a decomposed UDT object will be
  transformed to the start offset (index of the first byte of its raw data) of
  its containing byte array. Every field accessing bytecode of this object will
  be transformed to the array accessing code based on the absolute field offset
  ({\ttfamily \small object\_start\_offset + relative\_field\_offset}). The
  offset computation depends on the raw data size of each UDT instance. In each
  SUDT, Deca synthesizes {\ttfamily \small static} fields or methods to offer
  data-size values of all the UDT fields. The data sizes of the primitive type fields
  are already defined in the official JVM specification, while the sizes of
  non-primitive type fields can be recursively got from the SUDTs of these
  fields. If a field data-size value can be determined, then it
  is stored as a global constant value in a static field in the
  corresponding SUDT. Otherwise, Deca synthesizes a static method of the SUDT
  to compute the data size during data processing.  Similarly, for each UDT,
  Deca synthesizes {\ttfamily \small static} fields or methods to offer
  relative offset values of all UDT fields in the SUDT. The relative offsets can
  be computed based on the field data sizes and the field order in the UDT
  definition. As an optimization technique, we can reorder the UDT fields by
  putting the fields with determinable sizes in the front. This method makes
  more field offset values can be determined by Deca. Once all the field
  offsets can be computed, Deca transforms every methods of the UDT into its
  SUDT. During the transformation, all the field accessing code are replaced
  with the array accessing code.

\item[Linking.] Deca only transforms the code of UDTs in the decomposition
  sub-phase. In this sub-phase, Deca processes the code of UDFs and the main
  method for each stage. Deca memory manager allocates byte arrays rather than
  object arrays for data caching and shuffling. In Spark, every task processes
  data objects sequentially in each UDT object array in a loop.  Because the
  array type is changed, the array index values in the loop must be computed
  based on the raw data sizes of each object in the optimized code. In fact,
  the array index variable stores the value of the start offset of the current
  processing data object. Base on the new array index values, Deca changes the
  code of UDFs and the main method in the following ways: 1) removes the
  invocations of UDT object constructors and directly write the initial values
  in the constructor parameters into the byte arrays based on the absolute field
  offset; 2) replaces all the field accessing code with the array accessing code;
  3) replaces each invocation of a UDT method with the corresponding SUDT method,
  and adds the byte array and the object start offset as the additional parameters
  of the invocation.

\item[Optimization.] After decomposition and linking, the Deca optimizer
  further performs classical program optimizations, such as
  constant/copy propagation, method inlining and loop fusion. Most JVM JIT
  compilers prefer smaller methods, but we use method inlining aggressively in
  the pre-processing and transformation phases. If we produce giant methods that
  have sizes exceeding a threshold value, we split each giant method into several
  smaller ones using program slicing techniques.

\end{description}

\lstset{frame=tb,
language=scala,
xleftmargin=3.5em,
aboveskip=1mm,
belowskip=0mm,
showstringspaces=false,
columns=flexible,
basicstyle={\linespread{0.8}\fontsize{7}{8}\ttfamily},
numbers=left,
numberstyle=\fontsize{8}{10}\color{gray},
keywordstyle=\color{blue},
keywordstyle=[3]\color{magenta},
commentstyle=\color{dkgreen},
stringstyle=\color{mauve},
frame=none,
breaklines=true,
breakatwhitespace=true,
tabsize=3
}

\begin{figure}[t]
\centering
\begin{lstlisting}
def computeGradient() =  { 
  val result = new Array[Double](D)
  var offset = 0
  while(offset < block.size) {
    var factor = 0.0
    val label = block.readDouble(offset)
    offset += 8
    for (i <- 0 to D) {
      val feature = block.readDouble(offset)
      factor += weights(i) * feature
      offset += 8
    }
    factor = (1 / (1 + exp(-label * factor)) - 1) * label
    offset -= 8 * D
    for (i <- 0 to D) {
      val value = block.readDouble(offset)
      result(i) = result(i) + feature * factor
      offset += 8
    }
  }
  result
}
\end{lstlisting}
\caption{The transformed code of Logistic Regression.}
\label{code:lr_trans}
\end{figure}

While the code produced by Deca are JVM bytecode, in Figure~\ref{code:lr_trans},
we provide the equivalent Scala code to illustrate the logic of the transformed LR
code. The illustrated part is corresponding to the gradient computation of the
original code in line 21-24 of Figure~\ref{code:lr}.

\section{Disk Swapping/Spilling}
\label{sec:swap}

When the working set size is larger than the available memory space of an executor,
Spark moves part of its data out of the memory. For cached RDDs, Spark uses the LRU
strategy to select the cache blocks for eviction. The evicted data will be directly
discarded or swapped to the local disk according to the user-specified
\textit{storage-level}. For shuffles, Spark always spills the partial data
into temporary files, and merges them into final files at the end of task executions.

For cached RDDs, Deca modifies the original LRU strategy to evict page
groups rather than cache blocks. Accessing in-page data through either page-infos or
pointers will refresh the corresponding page group's recently-used counter. Spark
serializes cache block data before write them into disk files, or transfer them
through network for non-local accesses. In Deca, the decomposed data bytes can be
directly used for disk and network I/O.

For shuffles, like Spark, Deca sorts the pointers before spilling, and writes the
spilled data into files according to the order of the pointers. If a shuffle buffer has only
pointers that reference page segments, Deca does not spill these pointers because
normally they only occupy a small memory space. It pauses the shuffling and triggers
the cache block eviction to make enough room. Deca uses a small memory space (normally only
one page) to merge sorted spilled files. Once the merging space is fully filled, the
merged data will be flushed to the final output file.

\end{appendix}

\end{document}